%% file: pldi.tex
\newif\ifdraft\draftfalse 
\newif\ifextended\extendedtrue 
\newif\ifextrafries\extrafriesfalse

\PassOptionsToPackage{final}{microtype} 
\ifextended
  \documentclass[sigplan,screen,authorversion]{acmart}
\else
  \documentclass[sigplan,screen]{acmart}
\fi
\usepackage{local}

\pretocmd{\proof}{%
  \setlength\parindent{0pt}%
  \setlength\parskip{1em}%
}{}{\typeout{Patching proof failed.}}

\begin{document}

\title{Scalable Verification of Probabilistic Networks}
\ifextended
  \titlenote{Extended version with appendix.}
  \def\authornotenum{2}
\else
  \def\authornotenum{1}
\fi

\settopmatter{authorsperrow=4}

\author{Steffen Smolka}
\affiliation{Cornell University\\Ithaca, NY, USA}

\author{Praveen Kumar}
\affiliation{Cornell University\\Ithaca, NY, USA}

\author{David M Kahn}
\affiliation{Carnegie Mellon University\\Pittsburgh, PA, USA}
\authornote{Work performed at Cornell University.}

\author{Nate Foster}
\affiliation{Cornell University\\Ithaca, NY, USA}

\author{Justin Hsu}
\affiliation{University of Wisconsin\\Madison, WI, USA}
\authornotemark[\authornotenum]

\author{Dexter Kozen}
\affiliation{Cornell University\\Ithaca, NY, USA}

\author{Alexandra Silva}
\affiliation{University College London\\London, UK}

\begin{abstract}
This paper presents \mcnetkat, a scalable tool for verifying
probabilistic network programs. \mcnetkat is based on a new semantics
for the guarded and history-free fragment of Probabilistic \netkat in
terms of finite-state, absorbing Markov chains. This view allows the
semantics of all programs to be computed exactly, enabling
construction of an automatic verification tool. Domain-specific
optimizations and a parallelizing backend enable
\mcnetkat to analyze networks with thousands of nodes,
automatically reasoning about general properties such as probabilistic
program equivalence and refinement, as well as networking properties
such as resilience to failures. We evaluate \mcnetkat's scalability
using real-world topologies, compare its performance against
state-of-the-art tools, and develop an extended case study on a
recently proposed data center network design.
\end{abstract}

\ccsdesc[500]{Theory of computation~Automated reasoning}
\ccsdesc[500]{Theory of computation~Program semantics}
\ccsdesc[100]{Theory of computation~Random walks and Markov chains}
\ccsdesc[300]{Networks~Network properties}
\ccsdesc[300]{Software and its engineering~Domain specific languages}

\keywords{Network verification, Probabilistic Programming}

\setcopyright{acmlicensed}
\acmPrice{15.00}
\acmDOI{10.1145/3314221.3314639}
\acmYear{2019}
\copyrightyear{2019}
\acmISBN{978-1-4503-6712-7/19/06}
\acmConference[PLDI '19]{Proceedings of the 40th ACM SIGPLAN Conference on Programming Language Design and Implementation}{June 22--26, 2019}{Phoenix, AZ, USA}
\acmBooktitle{Proceedings of the 40th ACM SIGPLAN Conference on Programming Language Design and Implementation (PLDI '19), June 22--26, 2019, Phoenix, AZ, USA}

\maketitle

\renewcommand{\shortauthors}{S. Smolka, P. Kumar, D. Kahn, N. Foster, J. Hsu, D. Kozen, and A. Silva}

\section{Introduction}
\label{sec:intro}

Networks are among the most complex and critical computing systems
used today. Researchers have long sought to develop automated
techniques for modeling and analyzing network behavior~\citep{xie},
but only over the last decade has programming language methodology
been brought to bear on the problem~\citep{openflow,sdn-cacm,p4},
opening up new avenues for reasoning about networks in a rigorous and
principled way~\citep{veriflow,hsa,AFGJKSW13a,FKMST15a,p4v}. Building
on these initial advances, researchers have begun to target more
sophisticated networks that exhibit richer phenomena. In particular,
there is renewed interest in \emph{randomization} as a tool for
designing protocols and modeling behaviors that arise in large-scale
systems---from uncertainty about the inputs, to expected load, to
likelihood of device and link failures.

Although programming languages for describing randomized networks
exist~\citep{probnetkat-cantor,bayonet}, support for automated
reasoning remains limited. Even basic properties require quantitative
reasoning in the probabilistic setting, and seemingly simple programs
can generate complex distributions. Whereas state-of-the-art tools can
easily handle deterministic networks with hundreds of thousands of
nodes, probabilistic tools are currently orders of magnitude behind.

This paper presents \mcnetkat, a new tool for reasoning about
probabilistic network programs written in the guarded and history-free
fragment of Probabilistic \netkat
(\probnetkat)~\citep{AFGJKSW13a,FKMST15a,probnetkat-scott,
probnetkat-cantor}. \probnetkat is an expressive programming language
based on Kleene Algebra with Tests, capable of modeling a variety of
probabilistic behaviors and properties including randomized
routing~\citep{valiant82,smore}, uncertainty about
demands~\citep{roy15}, and failures~\citep{gill11}. The history-free
fragment restricts the language semantics to input-output behavior
rather than tracking paths, and the guarded fragment provides
conditionals and while loops rather than union and iteration
operators. Although the fragment we consider is a restriction of the
full language, it is still expressive enough to encode a wide range of
practical networking models. Existing deterministic tools, such as
Anteater~\citep{anteater}, HSA~\citep{hsa}, and
Veriflow~\citep{veriflow}, also use guarded and history-free models.

To enable automated reasoning, we first reformulate the semantics
of \probnetkat in terms of finite state Markov chains. We introduce
a \emph{big-step} semantics that models programs as Markov chains that
transition from input to output in a single step, using an
auxiliary \emph{small-step} semantics to compute the closed-form
solution for the semantics of the iteration operator. We prove that
the Markov chain semantics coincides with the domain-theoretic
semantics for \probnetkat developed in previous
work~\citep{probnetkat-scott, probnetkat-cantor}. Our new
semantics also has a key benefit: the limiting distribution of the resulting
Markov chains can be computed exactly in closed form, yielding a concise
representation that can be used as the basis for building a
practical tool.

We have implemented \mcnetkat in an OCaml prototype that takes a
\probnetkat program as input and produces a stochastic matrix
that models its semantics in a finite and explicit form. \mcnetkat
uses the UMFPACK linear algebra library as a back-end solver to
efficiently compute limiting distributions~\citep{UMFPACK}, and
exploits algebraic properties to automatically parallelize the
computation across multiple machines. To facilitate comparisons with
other tools, we also developed a back-end based on
PRISM~\citep{kwiatkowska2011prism}.

To evaluate the scalability of \mcnetkat, we conducted experiments on
realistic topologies, routing schemes, and properties. Our results
show that \mcnetkat scales to networks with thousands of switches, and
performs orders of magnitude better than a state-of-the-art tool based
on general-purpose symbolic inference~\citep{bayonet,psi}. We also
used \mcnetkat to carry out a case study of the resilience of
a fault-tolerant data center design proposed by~\citet{liu2013f10}.

\paragraph*{Contributions and outline.}
The central contribution of this paper is the development of a {\em
scalable probabilistic network verification tool}. We develop a new,
tractable semantics that is sound with
respect to \probnetkat's original denotational model. We present a
prototype implementation and evaluate it on a variety of scenarios
drawn from real-world networks. In \cref{sec:overview},
we introduce \probnetkat using a
running example. In \cref{sec:big-step}, we present a semantics based
on {\em finite stochastic matrices} and show that it fully
characterizes the behavior of \probnetkat programs
(\cref{thm:big-step-sound}). In \cref{sec:small-step}, we show how to
compute the matrix associated with iteration in closed form.
In \cref{sec:implementation}, we discuss our implementation,
including symbolic data structures and optimizations that are needed
to handle the large state space efficiently. In \cref{sec:experiments},
we evaluate the scalability of \mcnetkat on a common data center
design and compare its performance against state-of-the-art probabilistic 
tools. In \cref{sec:case}, we present a case study
using \mcnetkat to analyze resilience in the presence of link failures. We survey related
work in \cref{sec:rw} and conclude in \cref{sec:conc}.
\ifextended
We defer proofs to the appendix.
\else
Proofs can be found in the extended version of this paper \cite{mcnetkat-full}.
\fi

\section{Overview}
\label{sec:overview}

\begin{figure}
\centerline{\includegraphics[scale=0.09,draft=false]{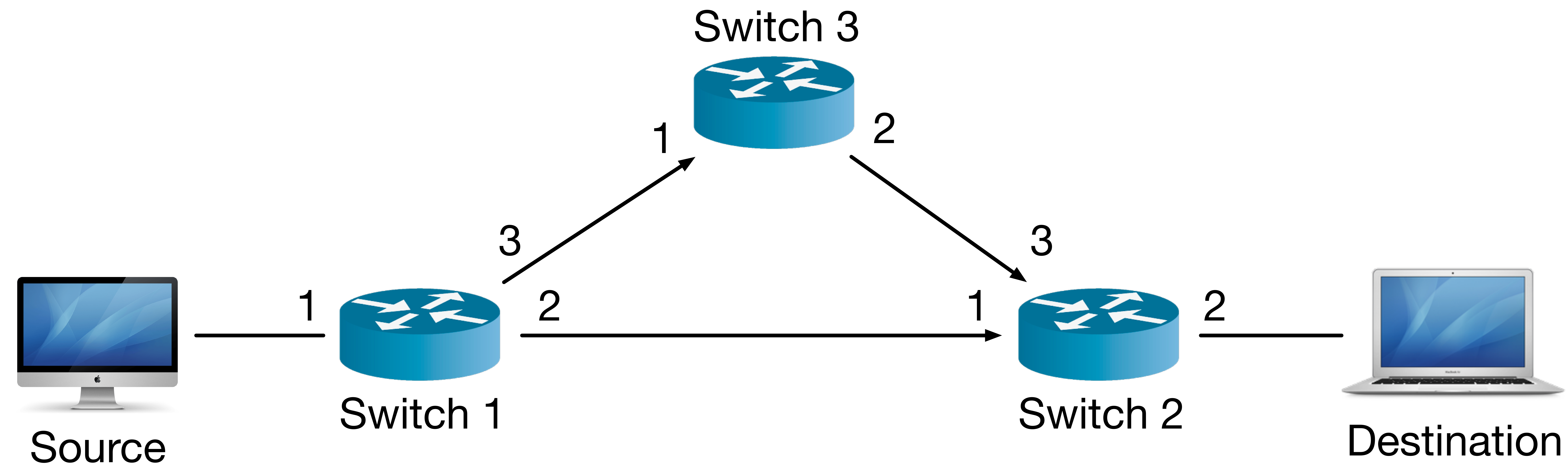}}
\vspace{-0.5em}
\caption{Network topology for running example.}
\label{fig:example-network}
\end{figure}

\newcommand\up[1]{\header{up}_{#1}}
\newcommand\cond[3]{\pseq{\match{#1}{#2}}{#3}}

This section introduces a running example that illustrates the main
features of the \probnetkat language as well as some quantitative
network properties that arise in practice.

\paragraph*{Background on \probnetkat.}
Consider the network in \Cref{fig:example-network}, which
connects a source to a destination in a topology with three switches.
We will first introduce a program that forwards packets from
the source to the destination, and then verify that it correctly
implements the desired behavior. Next, we will show how to enrich our program to model the
possibility of link failures, and develop a fault-tolerant forwarding
scheme that automatically routes around failures. Using a quantitative
version of program refinement, we will show that the fault-tolerant
program is indeed more resilient than the initial program. Finally, we
will show how to compute the expected degree of resilience analytically.


To a first approximation, a \probnetkat program can be thought of as a
randomized function that maps input packets to sets of output packets.
Packets are modeled as records, with fields for standard
headers---such as the source (\src) and destination (\dst)
addresses---as well as two fields switch (\sw) and port (\pt) encoding
the current location of the packet. \probnetkat provides several
primitives for manipulating packets: a \emph{modification}
$\modify{\field}{n}$ returns the input packet with field $\field$
updated to $n$, while a \emph{test} $\match{\field}{n}$ returns either
the input packet unmodified if the test succeeds, or the empty set if
the test fails. The primitives $\skp$ and $\drp$ behave like a test
that always succeeds and fails, respectively. In the guarded fragment
of the language, programs can be composed sequentially
($\pseq{\polp}{\polq}$), using conditionals
($\ite{\polp}{\polq_1}{\polq_2}$), while loops
($\while{\polp}{\polq}$), or probabilistic choice
($\polp \oplus \polq$).

Although \probnetkat programs can be freely constructed by composing
primitive operations, a typical network model is expressed using two
programs: a \emph{forwarding} program (sometimes called a \emph{policy}) and a
\emph{link} program (sometimes called a \emph{topology}). The forwarding
program describes how packets are transformed locally by the switches
at each hop. In our running example, to route packets from the source
to the destination, switches $1$ and $2$ can simply forward all
incoming packets out on port $2$ by modifying the port field ($\pt$).
This program can be encoded in \probnetkat by performing a case
analysis on the location of the input packet, and then setting the
port field to $2$:
\begin{align*}
\polp \defeq{}
&\ite{\match{\sw}{1}}{\modify{\pt}{2}}{}\\
&\ite{\match{\sw}{2}}{\modify{\pt}{2}}{\drop}
\end{align*}
The final $\drop$ at the end of this program encodes the policy for
switch 3, which is unreachable.

We can model the topology as a cascade of conditionals that match
packets at the end of each link and update their locations to the
link's destination:
\begin{align*}
t &\defeq
\ite{\pseq{\match{\sw}{1}}{\match{\pt}{2}}}{\pseq{\modify{\sw}{2}}{\modify{\pt}{1}}}{\dots}
\end{align*}
To build the overall network model, we first define predicates for the
ingress and egress locations,
\begin{align*}
  \mathit{in} \defeq \pseq{\match{\sw}{1}}{\match{\pt}{1}} &&
  \mathit{out} \defeq \pseq{\match{\sw}{2}}{\match{\pt}{2}}
\end{align*}
and then combine the forwarding policy $\polp$ with the topology $t$.
More specifically, a packet traversing the network starts at an
ingress and is repeatedly processed by switches and links until it
reaches an egress:
\[
  \model(p,t) \defeq \pseq{in}{\pseq{\polp}{\while{\pnot{\mathit{out}}}{(\pseq{t}{\polp})}}}
\]
We can now state and prove properties about the network by reasoning
about this model. For instance, the following equivalence states that
$\polp$ forwards all packets to the destination:
\[
  \model(p,t) \ \equiv \ \pseq{in}{\pseq{\modify{\sw}{2}}{\modify{\pt}{2}}}
\]
The program on the right can be regarded as an ideal specification that
``teleports'' each packet to its destination. Such equations were also used in
previous work to reason about properties such as waypointing, reachability,
isolation, and loop freedom~\citep{AFGJKSW13a,FKMST15a}.

\paragraph*{Probabilistic reasoning.}
Real-world networks often ex\-hib\-it nondeterministic behaviors such as
fault tolerant routing schemes to handle unexpected
failures~\citep{liu2013f10} and randomized algorithms to balance
load across multiple paths~\citep{smore}. Verifying that networks
behave as expected in these more complicated scenarios requires a
form of probabilistic reasoning, but most state-of-the-art network
verification tools model only deterministic
behaviors~\citep{veriflow,hsa,FKMST15a}.

To illustrate, suppose we want to extend our example with
link failures. Most modern switches execute low-level
protocols such as Bidirectional Forwarding Detection (BFD) that
compute real-time health information about the link connected to each
physical port~\citep{rfc7130}. We can enrich our model so that each
switch has a boolean flag $\up{i}$ that indicates whether the
link connected to the switch at port $i$ is up. Then, we can adjust the
forwarding logic to use backup paths when the link is down: for switch $1$,
\begin{align*}
\hat{\polp}_1 \defeq{}
  &\ite{\match{\up{2}}{1}}{\modify{\pt}{2}}{}\\
  &\ite{\match{\up{2}}{0}}{\modify{\pt}{3}}{\drop}
\end{align*}
and similarly for switches $2$ and $3$. As before, we can package the
forwarding logic for all switches into a single program:
\begin{align*}
\hat{\polp} &\defeq
\ite{\match{\sw}{1}}{\hat{p_1}}{\ite{\match{\sw}{2}}{\hat{p_2}}{\hat{p_3}}}
\end{align*}
Next, we update the encoding of our topology to faithfully model link failures.
Links can fail for a wide variety of reasons, including human errors,
fiber cuts, and hardware faults. A natural way to model such failures
is with a \emph{probabilistic model}---i.e., with a distribution that
captures how often certain links fail:
\begin{align*}
f_0 &\defeq \pseq{
    \modify{\up{2}}{1}
  }{
    \modify{\up{3}}{1}
  }\\
f_1 &\defeq
  \oplus \big\{
    f_0 \withp  \tfrac{1}{2}, 
    (\pseq{ \modify{\up{2}}{0} } { \modify{\up{3}}{1} }) \withp \tfrac{1}{4},
    (\pseq{ \modify{\up{2}}{1} } { \modify{\up{3}}{0} }) \withp \tfrac{1}{4} \big\}\\
f_2 &\defeq \pseq{
    (\modify{\up{2}}{1} \oplus_{.8} \modify{\up{2}}{0})
  }{
    (\modify{\up{3}}{1} \oplus_{.8} \modify{\up{3}}{0})
  }
\end{align*}
Intuitively, in $f_0$ no links fail, in $f_1$ the links $\ell_{12}$
and $\ell_{13}$ fail with probability $25\%$ but at most one link
fails, while in $f_2$ the links fail independently with probability
$20\%$. Using the $\header{up}$ flags, we can model a topology with 
possibly faulty links like so:
\begin{align*}
\hat{t} &\defeq
\ite{\match{\sw}{1} \cmp \match{\pt}{2} \cmp \match{\up{2}}{1}}
    {\modify{\sw}{2} \cmp \modify{\pt}{1}}
    {\dots}
\end{align*}
Combining the policy, topology, and failure model yields a
model of the entire network:
\begin{align*}
\hat\model(p,t,f) \defeq
  ~&\kw{var}~\modify{\up{2}}{1}~\kw{in}\\
  &\kw{var}~\modify{\up{3}}{1}~\kw{in}\\
  &\model((\pseq{f}{p}), t)
\end{align*}
This refined model $\hat{\model}$ wraps our previous model $\model$
with declarations of the two local fields $\up{2}$ and $\up{3}$ and
executes the failure model ($f$) at each hop before executing the
programs for the switch ($p$) and topology ($t$).

Now we can analyze our resilient routing scheme $\hat{\polp}$. As a
sanity check, we can verify that it delivers packets to their
destinations in the absence of failures. Formally, it behaves like the
program that teleports packets to their destinations:
\[
  \hat{\model}(\hat{\polp}, \hat{t}, f_0) \ \equiv \
  \pseq{in}{\pseq{\modify{\sw}{2}}{\modify{\pt}{2}}}
\]
More interestingly, $\hat{\polp}$ is 1-resilient---i.e., it
delivers packets provided at most one link fails. Note that this
property does \emph{not} hold for the original, naive routing scheme
$\polp$:
\[
  \hat{\model}(\hat{\polp}, \hat{t}, f_1)
  \ \equiv \
  \pseq{in}{\pseq{\modify{\sw}{2}}{\modify{\pt}{2}}}
  \ \nequiv \
  \hat{\model}(\polp, \hat{t}, f_1)
\]
While $\hat{\polp}$ is not fully resilient under failure model $f_2$,
which allows two links to fail simultaneously, we can still
show that the refined routing scheme $\hat{\polp}$ performs strictly
better than the naive scheme $\polp$ by checking
\[
  \hat\model(p,\hat{t},f_2) \ < \ \hat\model(\hat{p},\hat{t},f_2)
\]
where $p < q$ intuitively means that $q$ delivers packets with higher
probability than $p$.

Going a step further, we might want to compute more general quantitative
properties of the distributions generated for a given program. For example, we
might compute the probability that each routing scheme delivers packets to the
destination under $f_2$ (i.e., $80\%$ for the naive scheme and $96\%$ for the
resilient scheme), potentially valuable information to help an Internet Service
Provider (ISP) evaluate a network design to check that it meets certain
service-level agreements (SLAs). With this motivation in mind, we aim to build
a scalable tool that can carry out automated reasoning on probabilistic network
programs expressed in \probnetkat.

\section{\probnetkat Syntax and Semantics}
\label{sec:probnetkat}

This section reviews the syntax of \probnetkat and presents a new
semantics based on finite state Markov chains.

\paragraph*{Preliminaries.}
A \emph{packet} $\pk$ is a record mapping a finite set of fields
$\field_1, \field_2, \dots, \field_k$ to bounded integers $n$. As we
saw in the previous section, fields can include standard header fields
such as source ($\src$) and destination ($\dst$) addresses, as well as
logical fields for modeling the current location of the packet in the
network or variables such as $\up{i}$. These logical fields are not
present in a physical network packet, but they can track auxiliary
information for the purposes of verification. We write $\pk.\field$ to
denote the value of field $\field$ of $\pi$ and $\upd{\pk}{\field}{n}$
for the packet obtained from $\pi$ by updating field $\field$ to hold
$n$. We let $\Pk$ denote the (finite) set of all packets.


\paragraph*{Syntax.}
\probnetkat terms can be divided into two classes: \emph{predicates}
($\preda,\predb,\ldots$) and \emph{programs} ($\polp,\polq,\ldots$).
Primitive predicates include \emph{tests} ($\match{\field}{n}$) and
the Boolean constants \emph{false} ($\pfalse$) and \emph{true}
($\ptrue$). Compound predicates are formed using the usual Boolean
connectives: disjunction ($\punion{\preda}{\predb}$), conjunction
($\pseq{\preda}{\predb}$), and negation ($\pnot{\preda}$). Primitive
programs include \emph{predicates} ($\preda$) and
\emph{assignments} ($\modify{\field}{n}$). The original version 
of the language also provides a $\pdup$ primitive, which logs the
current state of the packet, but the history-free fragment omits this
operation. Compound programs can be formed using \emph{parallel
composition} ($\punion{\polp}{\polq}$), \emph{sequential composition}
($\pseq{\polp}{\polq}$), and \emph{iteration} ($\pstar{\polp}$). In
addition, the \emph{probabilistic choice} operator $\polp \opr \polq$
executes $\polp$ with probability $r$ and $\polq$ with probability
$1-r$, where $r$ is rational, $0\le r\le 1$.
We sometimes use an $n$-ary version and omit the $r$'s:
$\polp_1\oplus\cdots\oplus\polp_n$
executes a $\polp_i$ chosen uniformly at random.
In addition to these core constructs (summarized
in \cref{fig:probnetkat}), many other useful constructs can be
derived. For example, mutable local variables (\eg, $\up{i}$, used to
track link health in \cref{sec:overview}), can be desugared into the language:
\begin{align*}
  \kw{var}~\modify{\field}{n}~\kw{in}~\polp \defeqs
  \pseq{\pseq{\modify{\field}{n}}{p}}{\modify{\field}{0}}
\end{align*}
Here $\field$ is a field that is local to $p$. The final assignment
$\modify{\field}{0}$ sets the value of $\field$ to a canonical value,
``erasing'' it after the field
goes out of scope. We often use local variables to record extra
information for verification---\eg, recording whether a packet
traversed a given switch allows reasoning about simple
waypointing and isolation properties, even though the history-free
fragment of \probnetkat does not model paths directly.

\paragraph{Guarded fragment.}
Conditionals and while loops can be encoded using union and iteration:
\begin{equation*}
\begin{aligned}
\ite{\preda}{\polp}{\polq}  &\defeqs
 \punion{\pseq{\preda}{\polp}}{\pseq{\pnot \preda}{\polq}}\\
\while{\preda}{\polp}       &\defeqs
 \pseq{(\pseq{\preda}{\polp})\star}{\pnot \preda}
\end{aligned}
\end{equation*}
Note that these constructs use the predicate $t$ as a \emph{guard},
resolving the inherent nondeterminism in the union and iteration
operators.
Our implementation handles programs in the guarded fragment of the
language---i.e., with loops and conditionals but without union and
iteration---though we will develop the theory in full generality here, to make
connections to previous work on \probnetkat clearer.  We believe this
restriction is acceptable from a practical perspective, as the main purpose of
union and iteration is to encode forwarding tables and network-wide processing,
and the guarded variants can often perform the same task. A notable exception is
multicast, which cannot be expressed in the guarded fragment.

\begin{figure}[t!]
\[
\begin{array}{r@{\kern1ex}r@{~}c@{~}l@{\kern2em}l}
\textrm{Naturals} & n & ::=  & \mathrlap{0 \mid 1 \mid 2 \mid \cdots}\\
\textrm{Fields} & \field & ::=  & \mathrlap{\field_1 \mid \cdots \mid \field_k} \\
\textrm{Packets} & \Pk \ni \pk & ::= & \mathrlap{\sset{\field_1=n_1, \dots , \field_k = n_k}} \\
\textrm{Probabilities} & r & \in & \mathrlap{[0,1] \cap \Q}\\[1ex]
\textrm{Predicates} & \preds &
   ::= & \pfalse                       & \textit{False} \\
    & & \mid & \ptrue                  & \textit{True} \\
    & & \mid & \match{\field}{n}       & \textit{Test} \\
    & & \mid & \punion{\preda}{\predb} & \textit{Disjunction} \\
    & & \mid & \pseq{\preda}{\predb}   & \textit{Conjunction} \\
    & & \mid & \pnot{\preda}           & \textit{Negation} \\[1ex]
\textrm{Programs} & \pols &
  ::= & \preda                         & \textit{Filter} \\
    & & \mid & \modify{\field}{n}      & \textit{Assignment} \\
    & & \mid & \punion{\polp}{\polq}   & \textit{Union} \\
    & & \mid & \pseq{\polp}{\polq}     & \textit{Sequence} \\
    & & \mid & \polp \opr \polq        & \textit{Choice} \\
    & & \mid & \pstar{\polp}           & \textit{Iteration}
\end{array}\]
\caption{\probnetkat Syntax.}
\label{fig:probnetkat}
\end{figure}

\paragraph*{Semantics.}
\label{sec:big-step}
Previous work on \probnetkat~\citep{probnetkat-cantor} modeled history-free
programs as maps $\pPk \to \Dist(\pPk)$, where $\Dist(\pPk)$ denotes the set of
probability distributions on $\pPk$. This semantics
is useful for establishing fundamental properties of the language, but we will
need a more explicit representation to build a practical verification tool.
Since the set of packets is finite, probability distributions over sets of
packets are discrete and can be characterized by a \emph{probability mass
function}, $f : \pPk \to [0,1]$ such that $\sum_{b \subseteq \Pk} f(b) = 1$. It
will be convenient to view $f$ as a \emph{stochastic vector} of
non-negative entries that sum to $1$.

A program, which maps inputs $a$ to distributions over outputs, can
then be represented by a square matrix indexed by $\Pk$ in which the
stochastic vector corresponding to input $a$ appears as the $a$-th
row. Thus, we can interpret a program $\polp$ as a matrix
$\bden{\polp} \in [0,1]^{\pPk\times\pPk}$ indexed by packet sets,
where the matrix entry $\bden{\polp}_{ab}$ gives the probability that
$\polp$ produces output $b \in \pPk$ on input $a \in \pPk$. The rows
of the matrix $\bden{\polp}$ are stochastic vectors, each encoding the
distribution produced for an input set $a$; such a matrix is
called \emph{right-stochastic}, or simply stochastic.
We write $\mathbb{S}(\pPk)$ for the set of right-stochastic matrices indexed
by $\pPk$.

\begin{figure}[t]
\raggedright\fbox{$\bden{\polp}~\in~\mathbb{S}(\pPk)$}\\
\begin{minipage}{.45\textwidth}
\begin{align*}
  \bden{\pfalse}_{ab} &\!\!\defeqs\!\! \ind{b = \emptyset}\\
  \bden{\ptrue}_{ab} &\!\!\defeqs\!\! \ind{a = b}\\
  \bden{\match{f}{n}}_{ab} &\!\!\defeqs\!\! \ind{b = \set{\pk \in a}{\pk.f = n}}\\
  \bden{\pnot{\preda}}_{ab} &\!\!\defeqs\!\! \ind{b \subseteq a} \cdot \bden{\preda}_{a,a-b}\\
  \bden{\modify{f}{n}}_{ab} &\!\!\defeqs\!\! \ind{b = \set{\pk[f:=n]}{\pk \in a}}\\
  \bden{\punion{\polp}{\polq}}_{ab} &\!\!\defeqs\!\!
    \sum_{c,d} \ind{c \cup d = b} \cdot \bden{\polp}_{a,c} \cdot \bden{\polq}_{a,d}\\
  \bden{\pseq{\polp}{\polq}} &\!\!\defeqs\!\! \bden{\polp} \cdot \bden{\polq}\\
  \bden{\polp \oplus_r \polq} &\!\!\defeqs\!\! r \cdot \bden{\polp} + (1-r) \cdot \bden{\polq}\\
  \bden{\polp\star}_{ab} &\!\!\defeqs\!\! \lim_{n \to \infty} \bden{p^{(n)}}_{ab}
\end{align*}
\end{minipage}
\caption{\probnetkat Semantics. The notation $\bden{\polp}_{ab}$ 
denotes the probability that $\polp$ produces $b$ on input $a$.}
\label{fig:big-step}
\end{figure}

\cref{fig:big-step} defines an interpretation of \probnetkat programs as
stochastic matrices; the Iverson bracket $\ind{\varphi}$ is $1$ if $\varphi$ is
true, and $0$ otherwise.  Deterministic program primitives are interpreted as
$\{0,1\}$-matrices---\eg, the program primitive $\pfalse$ is interpreted as the
following stochastic matrix:
\begin{align}
\label{eq:smatric=MC}
\setlength{\kbcolsep}{0pt}
\setlength{\kbrowsep}{0pt}
\setlength{\arraycolsep}{2pt}
\def\arraystretch{1.2}
\bden{\pfalse} =
\kbordermatrix{
          & \emptyset & b_2     & \ldots & b_n     \\
\emptyset & 1         & 0       & \cdots & 0       \\
\vdots    & \vdots    & \vdots  & \ddots & \vdots  \\
a_n       & 1         & 0       & \cdots & 0
}
&&&\quad
\begin{tikzpicture}[font=\footnotesize,baseline=-4ex]
\begin{scope}[execute at begin node=$, execute at end node=$, node distance=1.5em,
  every node/.style = {draw}
  ]
  \node (A2) {a_2};
  \node[below of=A2,draw=none] (A3) {\varvdots};
  \node[below of=A3] (AN) {a_n};
  \node[right of=A3, node distance=3.5em] (empty) {a_1 = \emptyset};
\end{scope}
\path[->,>=stealth,semithick]
  (A2) edge[bend left=30]                           node[above]{1} (empty)
  (AN) edge[bend right=30]                          node[below]{1} (empty)
  (empty) edge[loop right,looseness=4] node{1} (empty);
\end{tikzpicture}
\end{align}
which assigns all probability mass to the $\emptyset$-column. Similarly,
$\ptrue$ is interpreted as the identity matrix. Sequential composition can be
interpreted as matrix product,
\begin{align*}
\bden{\pseq{\polp}{\polq}}_{ab} &= \sum_{c} \bden{\polp}_{ac} \cdot \bden{\polq}_{cb}
= (\bden{\polp} \cdot \bden{\polq})_{ab}
\end{align*}
which reflects the intuitive semantics of composition: to step from
$a$ to $b$ in $\bden{\pseq{\polp}{\polq}}$, one must step from $a$ to
an intermediate state $c$ in $\bden{\polp}$, and then from $c$ to $b$ in
$\bden{\polq}$.

As the picture in \cref{eq:smatric=MC} suggests, a stochastic matrix
$B \in \mathbb{S}(\pPk)$ can be viewed as a \emph{Markov chain}
(MC)---i.e., a probabilistic transition system with state space
$\pPk$. The $B_{ab}$ entry gives the probability that the system
transitions from $a$ to $b$.

\paragraph*{Soundness.}
The matrix $\bden{\polp}$ is equivalent to the denotational semantics
$\den{\polp}$ defined in previous work~\citep{probnetkat-cantor}.
\begin{theorem}[Soundness]
\label{thm:big-step-sound}
Let $a,b \in \pPk$. The matrix
$\bden{\polp}$ satisfies $\bden{\polp}_{ab} = \den{\polp}(a)(\sset{b})$.
\end{theorem}
Hence, checking program equivalence for $\polp$ and $\polq$ reduces to
checking equality of the matrices $\bden{\polp}$ and $\bden{\polq}$.
\begin{corollary}
\label{cor:den-equal-iff-bden-equal}
$\den{\polp} = \den{\polq}$ if and only if $\bden{\polp} = \bden{\polq}$.
\end{corollary}
In particular, because the Markov chains are all finite state, the transition
matrices are finite dimensional with rational entries.  Accordingly, program
equivalence and other quantitative properties can be automatically verified
provided we can compute the matrices for given programs. This is relatively
straightforward for program constructs besides $\bden{\polp\star}$, whose matrix
is defined in terms of a limit. The next section presents a closed-form
definition of the stochastic matrix for this operator.

\section{Computing Stochastic Matrices}
\label{sec:small-step}

The semantics developed in the previous section can be viewed as a
``big-step'' semantics in which a single step models the execution of
a program from input to output. To compute the semantics of
$\polp\star$, we will introduce a finer, ``small-step'' chain in which a
transition models one iteration of the loop.

\begin{figure}
\input{chain-sketch}
\caption{The small-step semantics is given by a Markov
  chain with states 
  $\config{\textit{program},\textit{input set},\textit{output
  accumulator}}$. The three dashed arrows can be collapsed into the
  single solid arrow, rendering the program component superfluous.}
\label{fig:small-step}
\end{figure}
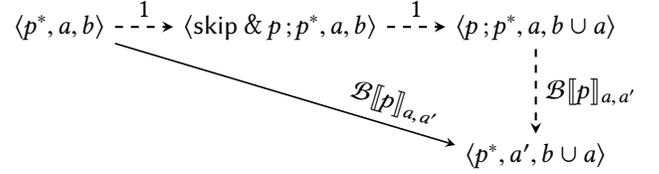

To build intuition, consider simulating $\polp\star$ using a
transition system with states given by triples $\config{\polp,a,b}$ in
which $\polp$ is the program being executed, $a$ is the set of (input)
packets, and $b$ is an accumulator that collects the output packets
generated so far. To model the execution of $\polp\star$ on input $a$,
we start from the initial state $\config{\polp\star,a,\emptyset}$ and
unroll $\polp\star$ one iteration according to the characteristic
equation $\polp\star \equiv \punion{\ptrue}{\pseq{\polp}{\polp\star}}$, 
yielding the following transition:
\begin{align*}
  \config{\polp\star,a,\emptyset} \stepsto{1}
  \config{\punion{\ptrue}{\pseq{\polp}{\polp\star}},a,\emptyset}
\end{align*}
Next, we execute both $\ptrue$ and $\pseq{\polp}{\polp\star}$ on the
input set and take the union of their results. Executing $\ptrue$ yields 
the input set as output, with probability 1:
\begin{align*}
  \config{\punion{\ptrue}{\pseq{\polp}{\polp\star}},a,\emptyset} \stepsto{1}
  \config{\pseq{\polp}{\polp\star},a,a}
\end{align*}
Executing $\pseq{\polp}{\polp\star}$, executes $\polp$ and feeds its
output into $\polp\star$:
\begin{align*}
  \forall a':\quad
  \config{\pseq{\polp}{\polp\star},a,a} \stepsto{\bden{\polp}_{a,a'}}
  \config{\polp\star,a',a}
\end{align*}
At this point we are back to executing $\polp\star$, albeit with a
different input set $a'$ and some accumulated output packets. The
resulting Markov chain is shown in \cref{fig:small-step}.

Note that as the first two steps of the chain are deterministic, we
can simplify the transition system by collapsing all three steps into
one, as illustrated in \cref{fig:small-step}. The program
component can then be dropped, as it now remains constant across
transitions. Hence, we work with a Markov chain over the state space
$\pPk \times \pPk$, defined formally as follows:
\begin{align*}
  \sden{\polp} &~\in~ \mathbb{S}(\pPk\times\pPk)\\
  \sden{\polp}_{(a,b),(a',b')} &\defeqs \ind{b' = b \cup a} \cdot \bden{\polp}_{a,a'} .
\end{align*}
We can verify that the matrix $\sden{\polp}$ defines a Markov
chain.
\begin{lemma}
\label{lem:small-step-stochastic}
$\sden{\polp}$ is stochastic.
\end{lemma}
Next, we show that each step in $\sden{\polp}$ models an iteration of
$\polp\star$. Formally, the $(n+1)$-step of $\sden{\polp}$ is
equivalent to the big-step behavior of the $n$-th unrolling of
$\polp\star$.
\begin{proposition}
\label{prop:small-step-characterization}
$\bden{\polp^{(n)}}_{a,b} = \sum_{a'} \sden{\polp}^{n+1}_{(a,\emptyset),(a',b)}$
\end{proposition}
Direct induction on the number of steps $n \geq 0$ fails because the
hypothesis is too weak. We generalize from start states with empty
accumulator to arbitrary start states.
\begin{lemma}\label{lem:star-mc-char}
  Let $p$ be program.
  Then for all $n \in \N$ and $a,b,b' \subseteq \Pk$, we have
  \[
  \sum_{a'} \ind{b' = a' \cup b} \cdot \bden{\polp^{(n)}}_{a,a'} =
  \sum_{a'} \sden{\polp}^{n+1}_{(a,b),(a',b')}
  .
  \]
\end{lemma}
\noindent
\cref{prop:small-step-characterization} then follows from
\cref{lem:star-mc-char} with $b=\emptyset$.

Intuitively, the long-run behavior of $\sden{\polp}$ approaches the
big-step behavior of $\polp\star$: letting $(a_n,b_n)$ denote the
random state of the Markov chain $\sden{\polp}$ after taking $n$ steps
starting from $(a,\emptyset)$, the distribution of $b_n$ for
$n\to\infty$ is precisely the distribution of outputs generated by
$\polp\star$ on input $a$ (by
\cref{prop:small-step-characterization} and the definition of
$\bden{\polp\star}$).

\label{sec:closed-form}
\paragraph*{Closed form.} 
The limiting behavior of finite state Markov chains has been
well studied in the literature (\eg, see \citet{kemeny1960finite}).
For so-called \emph{absorbing} Markov chains, the limit distribution
can be computed exactly. A state $s$ of a Markov chain $T$
is \emph{absorbing} if it transitions to itself with probability $1$,
\begin{align*}
\begin{tikzpicture}[->,>=stealth,shorten >=0.3pt,semithick,baseline=-.5ex]
\tikzstyle{every state}=[minimum size=1.5em]
\tikzset{every loop/.style={in=20,out=-20,looseness=8}}
\node[state] (s) {$s$};
\path (s) edge [loop right] node {$1$} (s);
\end{tikzpicture}
&&&\text{(formally: $T_{s,s\smash{'}} = \ind{s=s'}$)}
\end{align*}
and a Markov chain $T \in \mathbb{S}(S)$ is \emph{absorbing} if each state
can reach an absorbing state:
\begin{align*}
  \forall s \in S.\ \exists s' \in S, n \geq 0.\ T^n_{s,s'} > 0 \text{ and } T_{s',s'} = 1
\end{align*}
The non-absorbing states of an absorbing MC are
called \emph{transient}. Assume $T$ is absorbing with $n_t$ transient
states and $n_a$ absorbing states. After reordering the states so that
absorbing states appear first, $T$ has the form
\begin{align*}
  T = \begin{bmatrix}
    I & 0\\
    R & Q
  \end{bmatrix}
\end{align*}
where $I$ is the $n_a \times n_a$ identity matrix, $R$ is an
$n_t \times n_a$ matrix giving the probabilities of transient states
transitioning to absorbing states, and $Q$ is an $n_t \times n_t$
matrix specifying the probabilities of transitions between transient
states. Since absorbing states never transition to transient states by
definition, the upper right corner contains a $n_a \times n_t$ zero
matrix.

From any start state, a finite state absorbing MC always ends up in an
absorbing state eventually, \ie the limit
$T^\infty \defeq \lim_{n\to\infty} T^n$ exists and has the form
\begin{align*}
  T^\infty = \begin{bmatrix}
    I & 0\\
    A & 0
  \end{bmatrix}
\end{align*}
where the $n_t\times n_a$ matrix $A$ contains the
so-called \emph{absorption probabilities}. This matrix satisfies the
following equation:
\begin{align*}
  A = (I + Q + Q^2 + \dots)\,R
\end{align*}
Intuitively, to transition from a transient state to an absorbing
state, the MC can take an arbitrary number of steps between transient
states before taking a single---and final---step into an absorbing
state. The infinite sum $X \defeq \sum_{n \geq 0} Q^n$ satisfies $X =
I + QX$, and solving for $X$ yields
\begin{align}
  X = (I - Q)^{-1} \quad \text{and} \quad A = (I - Q)^{-1} R . \label{eq:inverse}
\end{align}
(We refer the reader to \citet{kemeny1960finite}
for the proof that the inverse exists.)

Before we apply this theory to the small-step semantics $\sden{-}$, it
will be useful to introduce some MC-specific notation. Let $T$ be an
MC. We write $s \reaches{T}_n s'$ if $s$ can reach $s'$ in precisely
$n$ steps, \ie if $T^n_{s,s'}>0$; and we write $s \reaches{T} s'$ if
$s$ can reach $s'$ in some number of steps, \ie if $T^n_{s,s'}>0$ for
some $n \geq 0$. Two states are said to \emph{communicate}, denoted
$s \communicate{T} s'$, if $s \reaches{T} s'$ and $s' \reaches{T} s$.
The relation $\communicate{T}$ is an equivalence relation, and its
equivalence classes are called \emph{communication classes}. A
communication class is \emph{absorbing} if it cannot reach any states
outside the class. Let $\prob{s \reaches{T}_n s'}$ denote the
probability $T^n_{s,s'}$. For the rest of the section, we fix a
program $\polp$ and abbreviate $\bden{\polp}$ as $B$ and
$\sden{\polp}$ as $S$. We also define {\em saturated states}, those
where the accumulator has stabilized.
\begin{definition}
\label{def:saturated-state}
A state $(a,b)$ of $S$ is called \emph{saturated} if $b$ has reached its
final value, \ie if $(a,b) \reaches{S} (a',b')$ implies $b'=b$.
\end{definition}
After reaching a saturated state, the output of $\polp\star$ is fully
determined. The probability of ending up in a saturated state with
accumulator $b$, starting from an initial state $(a,\emptyset)$, is
\begin{equation*}
  \lim_{n \to \infty} \sum_{a'} S^n_{(a,\emptyset),(a',b)}
\end{equation*}
and, indeed, this is the probability that $\polp\star$ outputs $b$ on input $a$ by
\cref{prop:small-step-characterization}.  Unfortunately, we cannot directly
compute this limit since saturated states are not necessarily
absorbing. To see this, consider $p\star =
(\modify{\field}{0} \oplus_{1/2}
\modify{\field}{1})\star $ over a single $\{ 0, 1 \}$-valued field $\field$. Then
$S$ has the form
\begin{center}
\begin{tikzpicture}[->,>=stealth,shorten >=0.5pt,auto,node distance=2cm,
                    semithick]
  \tikzstyle{every state}=[draw=none,text=black]

\matrix (M) [matrix of nodes, row sep=1ex, column sep=2.5em] {
                  & $0,0$ & $0,\sset{0,1}$\\
    $0,\emptyset$ &       &               \\
                  & $1,0$ & $1,\sset{0,1}$\\
};

  \path (M-2-1) edge                                        (M-1-2.west)
        (M-2-1) edge                                        (M-3-2.west)
        (M-1-2) edge                                        (M-3-2)
        (M-1-2) edge[loop right]                            (M-1-2)
        (M-3-2) edge                                        (M-3-3)
        (M-1-3) edge[<->]                                   (M-3-3)
        (M-1-3) edge[loop right,looseness=5.5,in=-11,out=11]  (M-1-3)
        (M-3-3) edge[loop right,looseness=5.5,in=-11,out=11]  (M-3-3);

  \draw (M-3-2) edge[shorten >=-2pt] (M-1-3.south west);
\end{tikzpicture}
\end{center}
where all edges are implicitly labeled with $\frac{1}{2}$, and $0$ and
$1$ denote the packets with $\field$ set to $0$ and $1$ respectively.
We omit states not reachable from $(0,\emptyset)$. The right-most
states are saturated, but they communicate and are thus not absorbing.

To align saturated and absorbing states, we can perform a quotient of this
Markov chain by collapsing the communicating states. We define an auxiliary
matrix,
\begin{align*}
  U_{(a,b),(a', b')} \defeq \ind{b'=b} \cdot \begin{cases}
  \ind{a'=\emptyset}
    &\text{if $(a,b)$ is saturated}  \\
  \ind{a'=a}
    &\text{else}
  \end{cases}
\end{align*}
which sends a saturated state $(a,b)$ to a canonical saturated state
$(\emptyset,b)$ and acts as the identity on all other states. In our
example, the modified chain $SU$ is as follows:
\begin{center}
\begin{tikzpicture}[->,>=stealth,shorten >=0.5pt,auto,node distance=2cm,
                    semithick]
  \tikzstyle{every state}=[draw=none,text=black]

\matrix (M) [matrix of nodes, row sep=1ex, column sep=2.5em] {
                  & $0,0$ & $0,\sset{0,1}$  &                       \\
    $0,\emptyset$ &       &                 & $\emptyset,\sset{0,1}$\\
                  & $1,0$ & $1,\sset{0,1}$  &                       \\
};

  \path (M-2-1) edge                                        (M-1-2.west)
        (M-2-1) edge                                        (M-3-2.west)
        (M-1-2) edge                                        (M-3-2)
        (M-1-2) edge[loop right]                              (M-1-2)
        (M-3-2) edge                                          (M-3-3)
        (M-1-3) edge                                     (M-2-4)
        (M-3-3) edge  (M-2-4)
        (M-2-4) edge[loop right,looseness=5.5,in=-11,out=11]  (M-2-4);

  \draw (M-3-2) edge[shorten >=-2pt] (M-1-3.south west);
\end{tikzpicture}
\end{center}
and indeed is absorbing, as desired. 
\begin{lemma}
\label{lem:monotone-chains}
  $S$, $U$, and $SU$ are \emph{monotone} in the sense that:
  $(a,b) \reaches{S} (a',b')$ implies $b \subseteq b'$ (and similarly
  for $U$ and $SU$).
\end{lemma}
\begin{proof*}
By definition ($S$ and $U$) and by composition ($SU$).\qedhere
\end{proof*}
\noindent
Next, we show that $SU$ is an absorbing MC:
\begin{proposition}
\label{prop:SU-properties}
Let $n \geq 1$.
\begin{enumerate}
\item \label{prop:SU-properties:(SU)*=S*U}
  $(SU)^n = S^nU$
\item \label{prop:SU-properties:absorbing}
$SU$ is an absorbing MC with absorbing states $\{(\emptyset, b)\}$.
\end{enumerate}
\end{proposition}
Arranging the states $(a,b)$ in lexicographically ascending order
according to $\subseteq$ and letting $n=|\pset\Pk|$, it then follows
from \cref{prop:SU-properties}.\ref{prop:SU-properties:absorbing} that
$SU$ has the form
\begin{equation*}
  SU = \begin{bmatrix}
    I_n & 0\\
    R              & Q
  \end{bmatrix}
\end{equation*}
where, for $a \neq \emptyset$, we have
\begin{equation*}
  (SU)_{(a,b),(a',b')} = \begin{bmatrix}R&Q\end{bmatrix}_{(a,b),(a',b')} .
\end{equation*}
Moreover, $SU$ converges and its limit is given by
\begin{equation}\label{eq:SU-limit-closed-form}
  (SU)^\infty \defeq \begin{bmatrix}
    I_n           & 0\\
    (I-Q)^{-1}R   & 0
  \end{bmatrix}
  = \lim_{n\to\infty} (SU)^n .
\end{equation}
Putting together the pieces, we can use the modified Markov chain $SU$ to
compute the limit of $S$.
\begin{theorem}[Closed Form]
\label{thm:closed-form}
Let $a,b,b' \subseteq \Pk$. Then
\[
  \lim_{n\to\infty} \sum_{a'} S^n_{(a,b),(a',b')} = (SU)^\infty_{(a,b),(\emptyset,b')} .
\]
%
The limit exists and can be computed exactly, in closed-form.
\end{theorem}

\section{Implementation}
\label{sec:implementation}

\begin{figure*}[t]
\input{img/pipeline}
\vspace{-0.5em}
\caption{Implementation using FDDs and a sparse linear algebra solver.}
\label{fig:compilation}
\end{figure*}
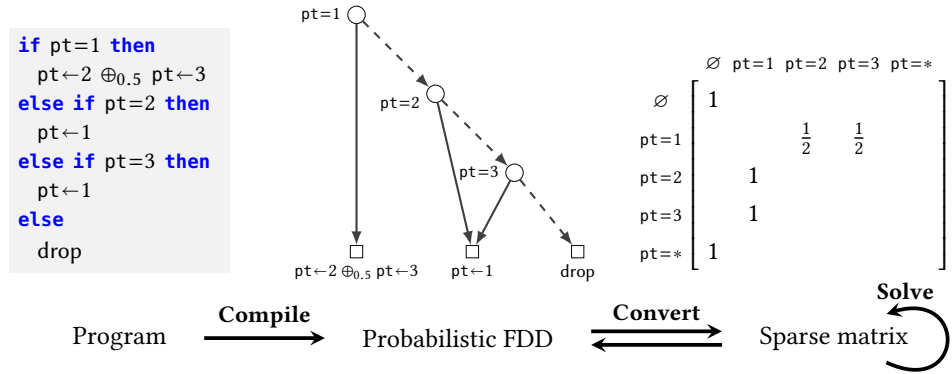

We have implemented \mcnetkat as an embedded DSL in OCaml in roughly
10KLoC.
The frontend provides functions for defining and manipulating
\probnetkat programs and for generating such programs automatically from
network topologies encoded using Graphviz.
These programs can then be analyzed by one of two backends:
the \emph{native backend} (PNK), which compiles programs to 
(symbolically represented) stochastic
matrices; or the \emph{PRISM-based backend} (PPNK), which emits inputs
for the state-of-the-art probabilistic model checker PRISM \citep{KwiatkowskaNP11}.

\paragraph*{Pragmatic restrictions.}
Although our semantics developed in \cref{sec:big-step} and
\cref{sec:small-step} theoretically supports computations on sets 
of packets, a direct implementation would be prohibitively
expensive---the matrices are indexed by the powerset $\pPk$ of the
universe of all possible packets! To obtain a practical analysis tool,
we restrict the state space to single packets. At the level of syntax,
we restrict to the guarded fragment of \probnetkat, \ie to programs
with conditionals and while loops, but without union and iteration.
This ensures that no proper packet sets are ever generated, thus
allowing us to work over an exponentially smaller state space. While
this restriction does rule out some uses of \probnetkat---most
notably, modeling multicast---we did not find this to be a serious
limitation because multicast is relatively uncommon in probabilistic
networking. If needed, multicast can often be modeled using multiple
unicast programs.





\subsection{Native Backend}

The native backend compiles a program to a symbolic representation of
its big step matrix. The translation, illustrated
in \Cref{fig:compilation}, proceeds as follows. First, we translate
atomic programs to Forwarding Decision Diagrams (FDDs), a symbolic
data structure based on Binary Decision Diagrams (BDDs) that encodes
sparse matrices compactly~\citep{compilekat}. Second, we translate
composite programs by first translating each sub-program to an FDD and
then merging the results using standard BDD algorithms. Loops require
special treatment: we (i) convert the FDD for the
body of the loop to a sparse stochastic matrix, (ii) compute the
semantics of the loop by using an optimized sparse linear
solver \cite{UMFPACK} to solve the system from \cref{sec:closed-form},
and finally (iii) convert the resulting matrix back to an FDD. We use
exact rational arithmetic in the frontend and FDD-backend to preempt
concerns about numerical precision, but trust the linear algebra
solver UMFPACK (based on 64 bit floats) to provide accurate
solutions.\footnote{UMFPACK is a mature library powering widely-used
scientific computing packages such as MATLAB and SciPy. } Our
implementation relies on several optimizations; we detail two of the
more interesting ones below.

\paragraph*{Probabilistic FDDs.}
Binary Decision Diagrams \citep{Akers:1978:BDD:1310167.1310815} and
variants thereof~\citep{Fujita:1997:MBD:607541.607565} have long been
used in verification and model checking to represent large state
spaces compactly. A variant called Forwarding Decision Diagrams (FDDs)
\citep{compilekat} was previously developed specifically for the networking
domain, but only supported deterministic behavior. In this work, we
extended FDDs to probabilistic FDDs. A probabilistic FDD is a rooted
directed acyclic graph that can be understood as a control-flow graph.
Interior nodes test packet fields and have outgoing true- and false-
branches, which we visualize by solid lines and dashed lines
in \Cref{fig:compilation}. Leaf nodes contain distributions
over \emph{actions}, where an action is either a set of modifications
or a special action $\pfalse$. To interpret an FDD, we start at the
root node with an initial packet and traverse the graph as dictated by
the tests until a leaf node is reached. Then, we apply each action in
the leaf node to the packet. Thus, an FDD represents a function of
type $\Pk \to \Dist(\Pk + \emptyset)$, or equivalently, a stochastic
matrix over the state space $\Pk + \emptyset$ where the
$\emptyset$-row puts all mass on $\emptyset$ by convention. Like BDDs,
FDDs respect a total order on tests and contain no isomorphic
subgraphs or redundant tests, which enables representing sparse
matrices compactly.

\paragraph*{Dynamic domain reduction.}
As \Cref{fig:compilation} shows, we do not have to represent the state
space $\Pk+\emptyset$ explicitly even when converting into sparse
matrix form. In the example, the state space is represented
by \emph{symbolic packets} $\pt = 1$, $\pt = 2$, $\pt = 3$, and $\pt =
*$, each representing an
\emph{equivalence class} of packets. For example,
$\pt = 1$ can represent all packets $\pk$ satisfying $\pk.\pt = 1$,
because the program treats all such packets in the same way. The
packet $\pt = *$ represents the set
$\set{\pk}{\pk.\pt \not\in \sset{1,2,3}}$. The symbol $*$ can be
thought of as a wildcard that ranges over all values not explicitly
represented by other symbolic packets. The symbolic packets are chosen
dynamically when converting an FDD to a matrix by traversing the FDD
and determining the set of values appearing in each field, either in a
test or a modification. Since FDDs never contain redundant tests or
modifications, these sets are typically of manageable size.

\subsection{PRISM backend}
\label{sec:prism-backend}

PRISM is a mature probabilistic model checker that has been actively
developed and improved for the last two decades. The tool takes as
input a Markov chain model specified symbolically in PRISM's input
language and a property specified using a logic such as Probabilistic
CTL, and outputs the probability that the model satisfies the
property. PRISM supports various types of models including finite
state Markov chains, and can thus be used as a backend for reasoning
about \probnetkat programs using our results from
\Cref{sec:big-step} and \Cref{sec:small-step}. Accordingly, we implemented a second
backend that translates \probnetkat to PRISM programs. While the
native backend computes the big step semantics of a program---a costly
operation that may involve solving linear systems to compute fixed
points---the PRISM backend is a purely syntactic transformation; the
heavy lifting is done by PRISM itself.

A PRISM program consists of a set of bounded variables together with a
set of transition rules of the form
\[
  \phi \ \rightarrow \ p_1 \cdot u_1 + \dots + p_k \cdot u_k
\]
where $\phi$ is a Boolean predicate over the variables, the $p_i$ are
probabilities that must sum up to one, and the $u_i$ are sequences of
variable updates. The predicates are required to be mutually exclusive
and exhaustive. Such a program encodes a Markov chain whose state
space is given by the finite set of variable assignments and whose
transitions are dictated by the rules: if $\phi$ is satisfied under
the current assignment $\sigma$ and $\sigma_i$ is obtained from
$\sigma$ by performing update $u_i$, then the probability of a transition from
$\sigma$ to $\sigma_i$ is $p_i$.

It is easy to see that any PRISM program can be expressed
in \probnetkat, but the reverse direction is slightly tricky: it
requires the introduction of an additional variable akin to a program
counter to emulate \probnetkat's control flow primitives such as loops
and sequences. As an additional challenge, we must be economical in
our allocation of the program counter, since the performance of model
checking is very sensitive to the size of the state space.

We address this challenge in three steps. First, we translate
the \probnetkat program to a finite state machine using a
Thompson-style construction \cite{Thompson68}.
Each edge is labeled with a predicate
$\phi$, a probability $p_i$, and an update $u_i$, subject to the
following well-formedness conditions:
\begin{enumerate}
  \item For each state, the predicates on its outgoing edges form a
  partition. 
\item For each state and predicate, the probabilities of
  all outgoing edges guarded by that predicate sum to one.
\end{enumerate}
Intuitively, the state machine encodes the control-flow graph.

This intuition serves as the inspiration for the next translation
step, which collapses each basic block of the graph into a single
state. This step is crucial for reducing the state space, since the
state space of the initial automaton is linear in the size of the program.
Finally, we obtain a PRISM program from the automaton as follows: for
each state $s$ with adjacent predicate $\phi$ and $\phi$-guarded
outgoing edges $s \xrightarrow{\phi/p_i/u_i} t_i$ for $1 \leq i \leq
k$, produce a PRISM rule
\[
  (\match{pc}{s} \land \phi) \ \rightarrow \
  p_1 \cdot (\pseq{u_1}\modify{pc}{t_1}) + \dots + p_k \cdot (\pseq{u_k}\modify{pc}{t_k}).
\] 
The well-formedness conditions of the state machine guarantee that the
resulting program is a valid PRISM program. With some care, the entire
translation can be implemented in linear time. Indeed, \mcnetkat
translates all programs in our evaluation to PRISM in under a second.

\section{Evaluation}
\label{sec:experiments}
\begin{figure}[t]
\input{img/fattree}
\caption{A FatTree topology with $\arity=4$.}
\label{fig:fattree}
\vspace{-0.5em}
\end{figure}
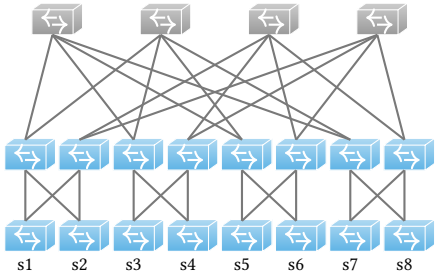

To evaluate \mcnetkat we conducted experiments on several benchmarks
including a family of real-world data center topologies and a
synthetic benchmark drawn from the literature \citep{bayonet}. We
evaluated \mcnetkat's scalability, characterized the effect of
optimizations, and compared performance against other state-of-the-art
tools. All \mcnetkat running times we report refer to the time needed to compile
programs to FDDs; the cost of comparing FDDs for equivalence and ordering, or of
computing statistics of the encoded distributions, is negligible.
All experiments were performed on machines with 16-core, 2.6
GHz Intel Xeon E5-2650 processors with 64 GB of memory.

\paragraph*{Scalability on FatTree topologies.}
\begin{figure}[t]
\begin{center}
\includegraphics[width=\columnwidth,draft=false]{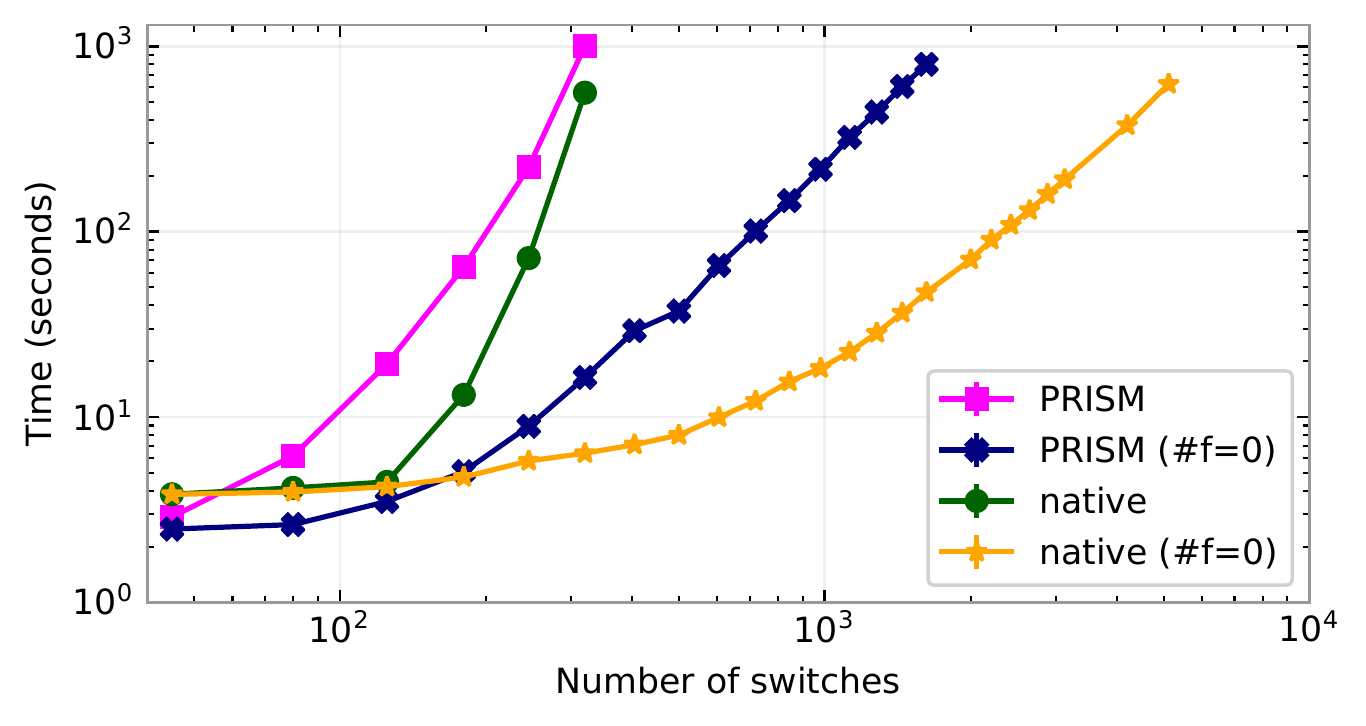}
\end{center}
\vspace{-1em}
\caption{Scalability on a family of data center topologies.}
\vspace{-0.5em}
\label{fig:scalability}
\end{figure}
We first measured the scalability of \mcnetkat by using it to compute
network models for a series of FatTree topologies of increasing size.
FatTrees~\citep{al2008scalable} (see also \Cref{fig:fattree}) are
multi-level, multi-rooted trees that are widely used as topologies
in modern data centers. FatTrees can be specified in
terms of a parameter $\arity$ corresponding to the number of ports on each
switch. A $\arity$-ary FatTree connects $\frac{1}{4}\arity^3$ servers using
$\frac{5}{4} \arity^2$ switches. To route packets, we used a form of
Equal-Cost Multipath Routing (ECMP) that randomly maps traffic flows
onto shortest paths. We measured the time needed to construct the
stochastic matrix representation of the program on a single machine
using two backends (native and PRISM) and under two failure
models (no failures and independent failures with probability
$1/1000$).

\Cref{fig:scalability} depicts the results, several of which  are worth discussing. First, the native backend scales quite well:
in the absence of failures ($f=0$), it scales to a
network with 5000 switches in approximately 10 minutes. This result
shows that \mcnetkat is able to handle networks of realistic size.
Second, the native backend consistently outperforms the PRISM backend.
We conjecture that the native backend is able to exploit algebraic
properties of the \probnetkat program to better parallelize the job.
Third, performance degrades in the presence of
failures. This is to be expected---failures lead to more complex probability
distributions which are nontrivial to represent and manipulate. 

\paragraph*{Parallel speedup.}
One of the contributors to \mcnetkat's good performance is its ability
to parallelize the computation of stochastic matrices across multiple
cores in a machine, or even across machines in a cluster. Intuitively,
because a network is a large collection of mostly independent devices,
it is possible to model its global behavior by first modeling the
behavior of each device in isolation, and then combining the results to
obtain a network-wide model. In addition to speeding up the
computation, this approach can also reduce memory usage,
often a bottleneck on large inputs.

To facilitate parallelization, we added an $n$-ary
disjoint branching construct to \probnetkat:
\begin{align*}
&\kw{case}~~\match{\sw}{1}~~\kw{then}~~p_1~~\kw{else}\\
&\kw{case}~~\match{\sw}{2}~~\kw{then}~~p_2~~\kw{else}\\
&\quad \dots\\
&\kw{case}~~\match{\sw}{n}~~\kw{then}~~p_n~~\phantom{else}
\end{align*}
Semantically, this construct is equivalent to a cascade of conditionals;
but the native backend compiles it in parallel using a map-reduce-style strategy,
using one process per core by default.

\begin{figure}
\begin{center}
\includegraphics[width=\columnwidth,draft=false]{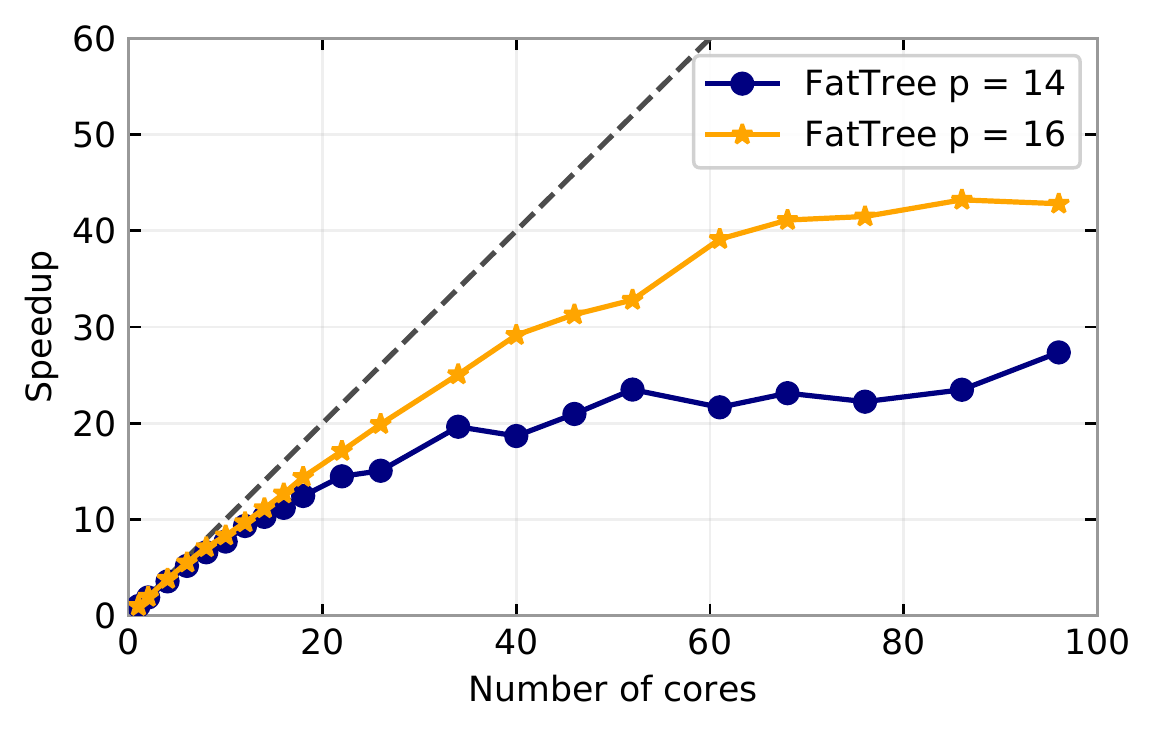}
\end{center}
\vspace{-1em}
\caption{Speedup due to parallelization.}
\vspace{-0.5em}
\label{fig:speedup}
\end{figure}

To evaluate the impact of parallelization, we compiled two
representative FatTree models ($\arity=14$ and $\arity=16$) using ECMP routing
on an increasing number of cores. With $m$ cores, we used one master
machine together with $r = \lceil{m/16 - 1}\rceil$ remote machines,
adding machines one by one as needed to obtain more physical cores.
The results are shown in \Cref{fig:speedup}. We see near linear speedup
on a single machine, cutting execution time by more than an order of
magnitude on our 16-core test machine. Beyond a single machine, the
speedup depends on the complexity of the submodels for each
switch---the longer it takes to generate the matrix for each switch,
the higher the speedup. For example, with a $\arity=16$ FatTree, we obtained
a 30x speedup using 40 cores across 3 machines.

\paragraph*{Comparison with other tools.}%
Bayonet~\citep{bayonet} is a state-of-the-art tool for analyzing
probabilistic networks.
Whereas \mcnetkat has a native backend tailored to the networking
domain and a backend based on a probabilistic model checker, Bayonet
programs are translated to a general-purpose probabilistic language
which is then analyzed by the symbolic inference engine
PSI~\citep{psi}. Bayonet's approach is more general, as it can model
queues, state, and multi-packet interactions under an asynchronous
scheduling model. It also supports Bayesian inference and parameter
synthesis. Moreover, Bayonet is fully symbolic whereas \mcnetkat uses
a numerical linear algebra solver \cite{UMFPACK} (based on floating
point arithmetic) to compute limits.

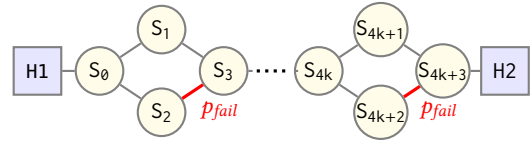
\begin{figure}[t]
\begin{tikzpicture}[>={Stealth[round]},auto,thick,scale=0.55,every node/.style = {scale=0.9}]
  \tikzstyle{host}=[rectangle,thick,draw=gray,fill=blue!10!white,minimum size=20pt,inner sep=0pt]
  \tikzstyle{switch}=[circle,thick,draw=gray,fill=yellow!10!white,minimum size=20pt,inner sep=0pt]

  \node[host]      (h1) at (0,0)      {\texttt{H1}};
  \node[switch]    (s0) at (1.5,0)    {$\mathtt{S_0}$};
  \node[switch]    (s1) at (3,1)    {$\mathtt{S_1}$};
  \node[switch]    (s2) at (3,-1)    {$\mathtt{S_2}$};
  \node[switch]    (s3) at (4.5,0)    {$\mathtt{S_3}$};

  \node[switch]    (sk0) at (6.8,0)         {$\mathtt{S_{4k}}$};
  \node[switch]    (sk1) at (8.3,1)    {$\mathtt{S_{4k+1}}$};
  \node[switch]    (sk2) at (8.3,-1)   {$\mathtt{S_{4k+2}}$};
  \node[switch]    (sk3) at (9.8,0)        {$\mathtt{S_{4k+3}}$};
  \node[host]      (h2)  at (11.3,0)      {$\mathtt{H2}$};

  \path[thick,gray] (h1) edge (s0);
  \path[thick,gray] (s0) edge (s1);
  \path[thick,gray] (s0) edge (s2);
  \path[thick,gray] (s1) edge (s3);
  \path[very thick,red] (s2) edge node[below right] {$p_{\textit{fail}}$} (s3);

  \path[thick,gray] (sk0) edge (sk1);
  \path[thick,gray] (sk0) edge (sk2);
  \path[thick,gray] (sk1) edge (sk3);
  \path[very thick,red] (sk2) edge node[below right] {$p_{\textit{fail}}$} (sk3);
  \path[thick,gray] (sk3) edge (h2);

  \draw[thick,gray] (s3.east) -- (5.2,0);
  \draw[thick,gray] (6.3,0) -- (sk0.west);
  \draw[very thick,black    ,dotted] (5.3,0) -- (6,0);
\end{tikzpicture}
\caption{Chain topology}
\vspace{-1em}
\label{fig:bayonet-topo}
\end{figure}

To evaluate how the performance of these approaches compares, we
reproduced an experiment from the Bayonet paper that analyzes the
reliability of a simple routing scheme in a family of ``chain''
topologies indexed by $k$, as shown in \cref{fig:bayonet-topo}.

For $k=1$, the network consists of four switches organized into a diamond,
with a single link that fails with probability $p_\textit{fail} =
1/1000$. For $k > 1$, the network consists of $k$ diamonds linked
together into a chain as shown in \cref{fig:bayonet-topo}. Within each
diamond, switch $S_0$ forwards packets with equal probability to
switches $S_1$ and $S_2$, which in turn forward to switch $S_3$.
However, $S_2$ drops the packet if the link to $S_3$ fails. We analyze
the probability that a packet originating at \texttt{H1} is
successfully delivered to \texttt{H2}. Our implementation does not
exploit the regularity of these topologies.

\begin{figure}[t]
\includegraphics[width=\columnwidth,draft=false,trim={5pt 5 0 -10},clip]{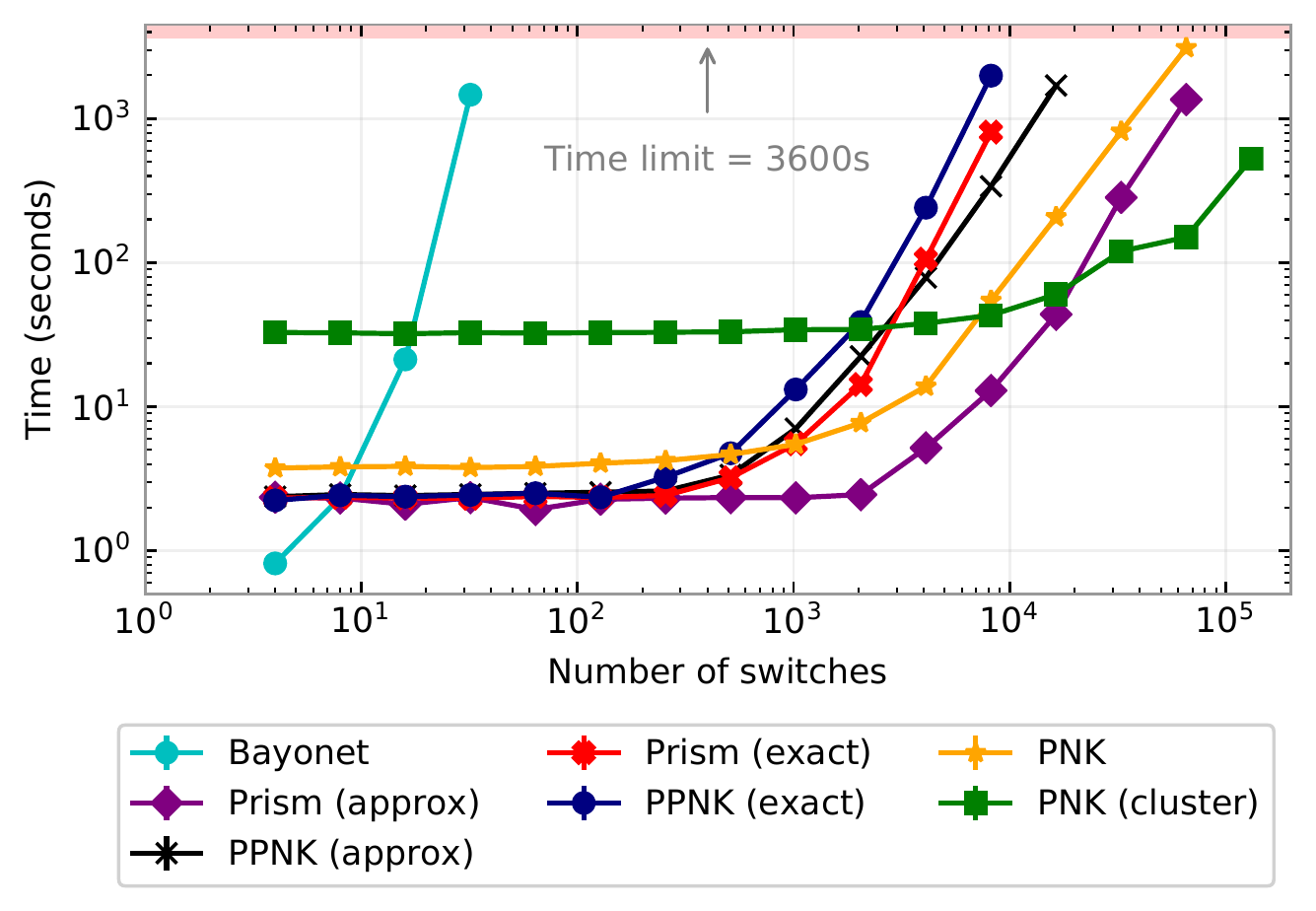}
\caption{Scalability on chain topology.}
\vspace{-1em}
\label{fig:bayonet}
\end{figure}
\Cref{fig:bayonet} gives the running time for several tools on this benchmark:
Bayonet, hand-written PRISM, \probnetkat with the PRISM backend
(PPNK), and \probnetkat with the native backend (PNK). Further, we ran
the PRISM tools in exact and approximate mode, and we ran
the \probnetkat backend on a single machine and on the cluster. Note
that both axes in the plot are log-scaled.

We see that Bayonet scales to 32 switches in about 25 minutes, before
hitting the one hour time limit and 64 GB memory limit at 48
switches. \probnetkat answers the same query for 2048 switches in
under 10 seconds and scales to over 65000 switches in about 50 minutes
on a single core, or just 2.5 minutes using a cluster of 24 machines.
PRISM scales similarly to \probnetkat, and performs best using the
hand-written model in approximate mode.

Overall, this experiment shows that for basic network verification tasks,
\probnetkat's domain-specific
backend based on specialized data structures and an optimized
linear-algebra library \cite{UMFPACK} can outperform an approach based
on a general-purpose solver.

\section{Case Study: Data Center Fault-Tolerance}
\label{sec:case}

\begin{figure*}[t!]
\begin{minipage}{.8\columnwidth}
\raggedleft
\input{img/abfattree}
\end{minipage}%
\begin{minipage}{.7\columnwidth}
\centering
\tabcolsep=0.11cm
\begin{tabular}{lccc}
\toprule
$k$ & \thead{$\hat\model(\ften{0}, f_k)$ \\ $\equiv$ \\${\mathit{teleport}}$}
    & \thead{$\hat\model(\ften{3}, f_k)$ \\ $ \equiv$ \\${\mathit{teleport}}$}
    & \thead{$\hat\model(\ften{3,5}, f_k)$ \\ $ \equiv$ \\${\mathit{teleport}}$} \\
\midrule
0 & \cmark & \cmark & \cmark \\
1 & \xmark & \cmark & \cmark \\
2 & \xmark & \cmark & \cmark \\
3 & \xmark & \xmark & \cmark \\
4 & \xmark & \xmark & \xmark \\
$\infty$ & \xmark & \xmark & \xmark \\
\bottomrule
\end{tabular}
\end{minipage}\hfill%
\begin{minipage}{.5\columnwidth}
\raggedright
\tabcolsep=0.11cm
\begin{tabular}{lccc}
\toprule
$k$ & \thead{$\mathtt{compare}$ \\ $\ften{0}$   \\ $\ften{3}$}
    & \thead{$\mathtt{compare}$ \\ $\ften{3}$   \\ $\ften{3,5}$}
    & \thead{$\mathtt{compare}$ \\ $\ften{3,5}$ \\ $\mathit{teleport}$}\\
\midrule
0 & $\equiv$ & $\equiv$ & $\equiv$ \\
1 & $<$ & $\equiv$ & $\equiv$ \\
2 & $<$ & $\equiv$ & $\equiv$ \\
3 & $<$ & $<$ & $\equiv$ \\
4 & $<$ & $<$ & $<$ \\
$\infty$ & $<$ & $<$ & $<$ \\
\bottomrule
\end{tabular}
\end{minipage}
\caption{%
  (a) AB FatTree topology with $\arity=4$.\hfill%
  (b) Evaluating $k$-resilience.\hfill%
  (c) Comparing schemes under $k$ failures.%
}
\label{fig:ften}
\end{figure*}
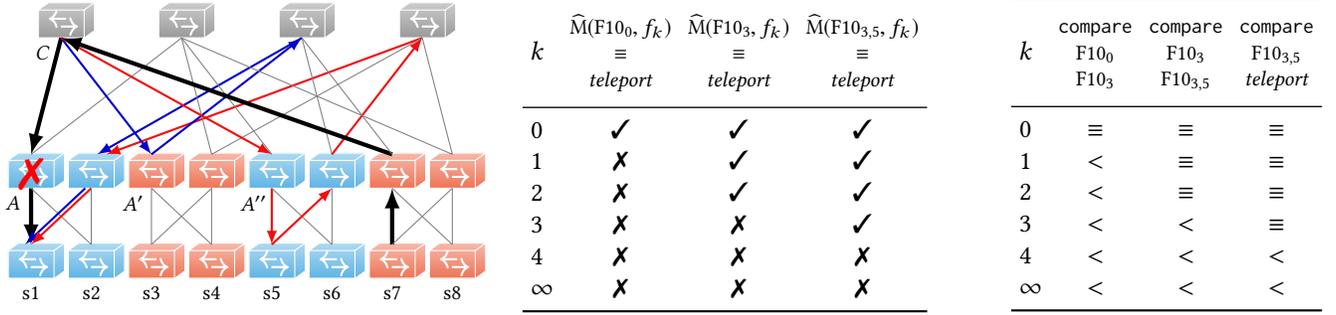


In this section, we go beyond benchmarks and present a case study that
illustrates the utility of \mcnetkat for probabilistic reasoning.
Specifically, we model the \ften{}~\citep{liu2013f10} data center
design in \probnetkat and verify its key properties.

\paragraph*{Data center resilience.}
An influential measurement study by~\citet{gill11} showed that data
centers experience frequent failures, which have a major impact on
application performance. To address this challenge, a number of
data center designs have been proposed that aim to simultaneously
achieve high throughput, low latency, and fault tolerance.

\paragraph*{\ften{} topology.}
\ften{} uses a novel topology called an \emph{AB FatTree},
see \Cref{fig:ften}(a), that enhances a
traditional FatTree~\citep{al2008scalable} with additional backup
paths that can be used when failures occur. To illustrate, consider
routing from $s_7$ to $s_1$ in \cref{fig:ften}(a) along one of the
shortest paths (in thick black). After reaching the core switch $C$ in a
standard FatTree (recall \Cref{fig:fattree}),
if the aggregation switch on the downward path
failed, we would need to take a 5-hop detour (shown in red) that goes
down to a different edge switch, up to a different core switch, and
finally down to $s_1$. In contrast, an AB FatTree~\citep{liu2013f10}
modifies the wiring of the aggregation later to provide shorter
detours---e.g., a 3-hop detour (shown in blue) for the previous
scenario.

\paragraph*{\ften{} routing.}%
\ften{}'s routing scheme uses three strategies to re-route packets after a
failure occurs. If a link on the current path fails and an equal-cost
path exists, the switch simply re-routes along that path. This
approach is also known as \emph{equal-cost multi-path routing} (ECMP).
If no shortest path exist, it uses a 3-hop detour if one is available,
and otherwise falls back to a 5-hop detour if necessary.

We implemented this routing scheme in \probnetkat in several steps.
The first, \ften{0}, approximates the hashing behavior of ECMP by
randomly selecting a port along one of the shortest paths to the
destination. The second, \ften{3}, improves the resilience of \ften{0}
by augmenting it with 3-hop re-routing---e.g., consider the blue path
in \cref{fig:ften}(a). We find a port on $C$ that connects to a
different aggregation switch $A'$ and forward the packet to $A'$. If
there are multiple such ports which have not failed, we choose one
uniformly at random. The third, \ften{3,5}, attempts 5-hop re-routing
in cases where $\ften{3}$ is unable to find a port on $C$ whose
adjacent link is up---e.g., consider the red path
in \cref{fig:ften}(a). The 5-hop rerouting strategy requires
a flag to distinguish packets taking a detour from regular packets.

\paragraph*{\ften{} network and failure model.}%
We model the network as discussed in \cref{sec:overview}, focusing on
packets destined to switch~1:
\[
  \model(p) \defeqs \pseq{in}{\dowhile{(\pseq{p}{t})}{(\pnot{\match{\sw}{1}})}}
\]
\mcnetkat automatically generates the topology program $t$ from
a Graphviz description. The ingress predicate $in$ is a disjunction of
switch-port tests over all ingress locations. Adding the failure model
and some setup code to declare local variables tracking the health of
individual links yields the complete network model:
\begin{align*}
\hat\model(p,f) \defeqs~
  &\kw{var}~\modify{\up{1}}{1}~\kw{in}~\dots~\kw{var}~\modify{\up{d}}{1}~\kw{in}~\model(\pseq{f}{p})
\end{align*}
Here, $d$ is the maximum degree of a topology node.
The entire model measures about 750 lines of \probnetkat code.

To evaluate the effect of different kinds of failures, we define a
family of failure models $f_k$ indexed by the maximum number of
failures $k \in \N \cup \sset{\infty}$ that may occur, where links fail
otherwise independently with probability $\mathit{pr}$; we leave
$\mathit{pr}$ implicit. To simplify the
analysis, we focus on failures occurring on downward paths (note
that \ften{0} is able to route around failures on the upward path,
unless the topology becomes disconnected).

\paragraph*{Verifying refinement.}
Having implemented \ften{} as a series of three refinements, we would
expect the probability of packet delivery to increase in each
refinement, but not to achieve perfect delivery in an unbounded
failure model $f_\infty$. Formally, we should have
\begin{align*}
 \drop < \hat\model(\ften{0},f_\infty) &< \hat\model(\ften{3},f_\infty) 
 \\&< \hat\model(\ften{3,5},f_\infty) <  {\mathit{teleport}}
\end{align*}
where $\mathit{teleport}$ moves the packet directly to its
destination, and $p < q$ means the probability assigned to every
input-output pair by $q$ is greater than the probability assigned by
$p$. We confirmed that these inequalities hold using \mcnetkat.


\begin{figure*}[t!]
\centering
\hspace*{\fill}
\includegraphics[width=.33\textwidth,draft=false]{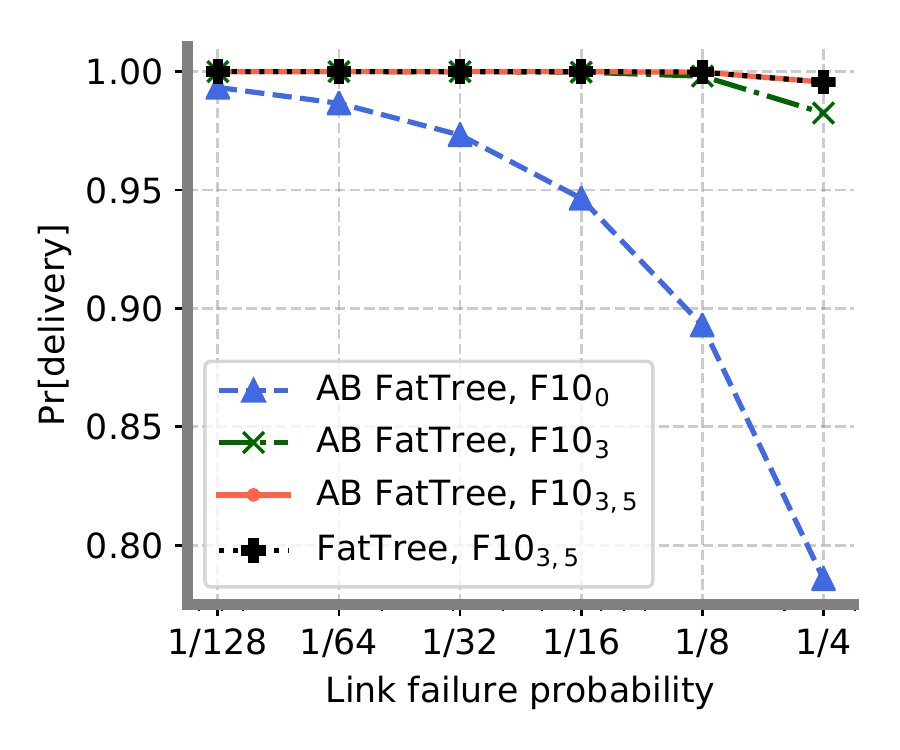}%
\hspace*{\fill}%
\includegraphics[width=.33\textwidth,draft=false]{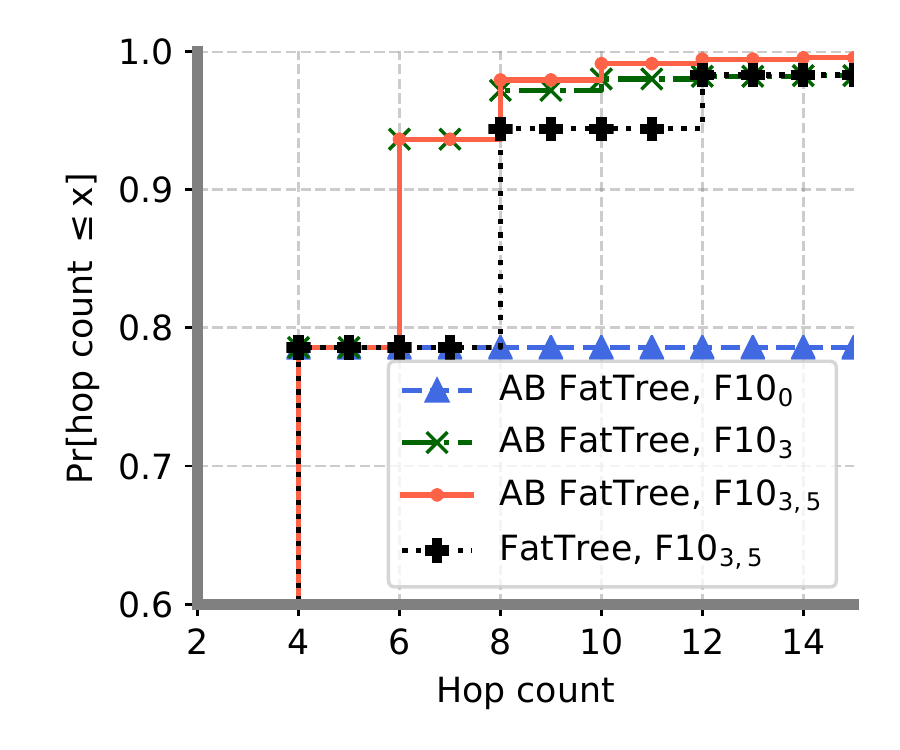}%
\hspace*{\fill}%
\includegraphics[width=.33\textwidth,draft=false]{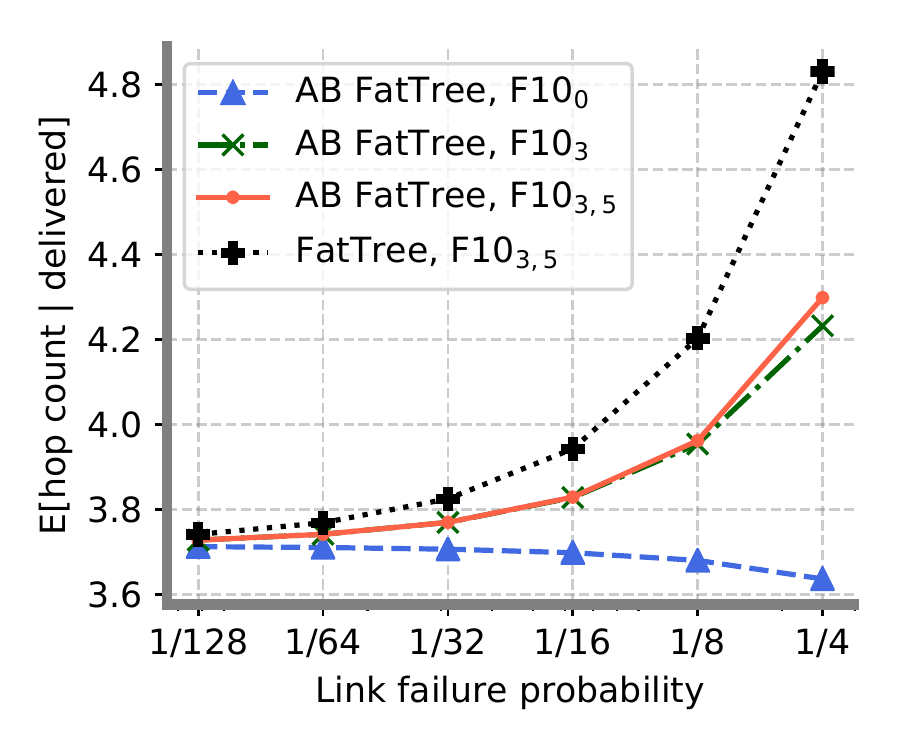}
\hspace*{\fill}
\\[-2em]
\null \footnotesize \hspace*{.18\textwidth} (a) \hfill \hspace*{1.5ex}(b) \hfill (c) \hspace*{.135\textwidth} \null
\caption{Case study results ($k=\infty$):
         (a) Probability of delivery vs. link-failure probability;
         (b) Increased path length due to resilience ($pr=1/4$);
         (c) Expected hop-count conditioned on delivery.}
\label{fig:case-studies}
\end{figure*}

\paragraph*{Verifying k-resilience.}
Resilience is the key property satisfied by \ften{}. By
using \mcnetkat, we were able to automatically verify that \ften{}
is resilient to up to three failures in the AB
FatTree~\cref{fig:ften}(a). To establish this property, we increased
the parameter $k$ in our failure model $f_k$ while checking
equivalence with teleportation (i.e., perfect delivery), as shown
in \cref{fig:ften}(b). The simplest scheme \ften{0} drops packets when
a failure occurs on the downward path, so it is 0-resilient. The
\ften{3} scheme routes around failures when a suitable aggregation switch 
is available, hence it is 2-resilient. Finally, the \ften{3,5} scheme
routes around failures as long as any aggregation switch is reachable,
hence it is 3-resilient. If the schemes are not equivalent to
$\mathit{teleport}$, we can still compare the relative resilience of
the schemes using the refinement order, as shown
in \cref{fig:ften}(c). Our implementation also enables precise,
quantitative comparisons. For example, \cref{fig:case-studies}(a)
considers a failure model in which an unbounded number of failures can
occur. We find that
\ften{0}'s delivery probability dips significantly as the failure probability
increases, while both \ften{3} and \ften{3,5} continue to ensure high
delivery probability by routing around failures.

\paragraph*{Analyzing path stretch.}
Routing schemes based on detours achieve a higher degree of resilience
at the cost of increasing the lengths of forwarding paths. We can
quantify this increase by augmenting our model with a counter that is
incremented at each hop and analyzing the expected path length.
\cref{fig:case-studies}(b) shows the cumulative distribution function of
latency as the fraction of traffic delivered within a given hop count.
On AB FatTree, \ften{0} delivers $\approx$80\% of the traffic in 4
hops, since the maximum length of a shortest path from any edge switch
to $s1$ is 4 and \ften{0} does not attempt to recover from
failures. \ften{3} and \ften{3,5} deliver the same amount of traffic
when limited to at most 4 hops, but they can deliver significantly
more traffic using $2$ additional hops by using 3-hop and 5-hop paths
to route around failures. \ften{3} also delivers more traffic with 8
hops---these are the cases when \ften{3} performs 3-hop re-routing
twice for a single packet as it encountered failure twice. We can also
show that on a standard FatTree, \ften{3,5} failures have a higher
impact on latency. Intuitively, the topology does not support 3-hop
re-routing. This finding supports a key claim of \ften{}: the topology
and routing scheme should be co-designed to avoid excessive path
stretch. Finally, \cref{fig:case-studies}(c) shows the expected path
length conditioned on delivery. As the failure probability increases,
the probability of delivery for packets routed via the core layer
decreases for \ften{0}. Thus, the distribution of delivered packets
shifts towards 2-hop paths via an aggregation switch, so the expected
hop-count decreases.



\section{Related Work}
\label{sec:rw}

The most closely related system to \mcnetkat is
Bayonet~\citep{bayonet}. In contrast to the domain-specific approach
followed in this paper, Bayonet uses a general-purpose probabilistic
programming language and inference tool~\citep{psi}. Such an approach,
which reuses existing techniques, is naturally appealing. In addition,
Bayonet is more expressive than \mcnetkat: it supports asynchronous
scheduling, stateful transformations, and probabilistic inference,
making it possible to model richer phenomena, such as congestion due
to packet-level interactions in queues. Of course, the extra
generality does not come for free. Bayonet requires programmers to
supply an upper bound on loops as the implementation is not guaranteed
to find a fixed point. As discussed
in \cref{sec:implementation}, \mcnetkat scales better than Bayonet on
simple benchmarks. Another issue is that writing a realistic scheduler
appears challenging, and one might also need to model host-level
congestion control protocols to obtain accurate results. Currently
Bayonet programs use deterministic or uniform schedulers and model
only a few packets at a time~\citep{bayonet-web}.

Prior work on ProbNetKAT \cite{probnetkat-scott} gave a measure-theoretic
semantics and an implementation that approximated programs using
sequences of monotonically improving estimates. While these estimates
were proven to converge in the limit,
\cite{probnetkat-scott}~offered no guarantees about the convergence rate. 
In fact, there are examples where the approximations do not converge
after any finite number of steps, which is obviously undesirable in a
tool. The implementation only scaled to 10s of switches. In contrast,
this paper presents a straightforward and implementable semantics; the
implementation computes limits precisely in closed form, and it scales
to real-world networks with thousands of switches.
\mcnetkat achieves this by restricting to the guarded and history-free fragment
of \probnetkat, sacrificing the ability to reason about
multicast and path-properties directly. In practice this sacrifice
seems well worth the payoff: multicast is somewhat
uncommon, and we can often reason about path-properties
by maintaining extra state in the packets.
In particular, \mcnetkat can still model the examples
studied in previous work by \citet{probnetkat-scott}.

Our work is the latest in a long line of
techniques using Markov chains as a tool for representing and
analyzing probabilistic programs. For an early example, see the
seminal paper of \citet{sharir1984verification}. Markov chains are
also used in many probabilistic model checkers, such as PRISM
\citep{kwiatkowska2011prism}.



Beyond networking applications, there are connections to other work on
verification of probabilistic programs. Di Pierro, Hankin,
and Wiklicky used probabilistic abstract interpretation to
statically analyze probabilistic
$\lambda$-calculus \citep{di2005probabilistic}; their work was
extended to a language $pWhile$, using a store and program
location state space similar to \citet{sharir1984verification}.
However, they do not deal with infinite limiting behavior beyond
stepwise iteration, and do not guarantee convergence. Olejnik,
Wicklicky, and Cheraghchi provided a probabilistic compiler $pwc$ for
a variation of $pWhile$
\citep{olejnik2016probabilistic}; their optimizations could potentially be useful for
\mcnetkat. A recent survey by \citet{gordon2014probabilistic} shows how to give
semantics for probabilistic processes using stationary distributions of
Markov chains, and studies convergence. Similar to our approach, they
use absorbing strongly connected components to represent termination. Finally,
probabilistic abstract interpretation is also an active area of
research~\citep{prob-AI}; it would be interesting to explore applications to
\probnetkat.

\section{Conclusion}
\label{sec:conc}

This paper presents a scalable tool for verifying probabilistic
networks based on a new semantics for the history-free fragment
of \probnetkat in terms of Markov chains. Natural directions for
future work include further optimization of our implementation---e.g.,
using Bayesian networks to represent joint distributions
compactly.
We are also interested in applying \mcnetkat to
other systems that implement algorithms for randomized
routing~\citep{smore,radwan18}, load balancing~\citep{dixit13},
traffic monitoring~\citep{csamp}, anonymity~\citep{tor}, and network
neutrality~\citep{neutrality14}, among others.

\begin{acks}
We are grateful to the anonymous reviewers and our shepherd Michael
Greenberg for their feedback and help in improving the paper. Thanks
also to Jonathan DiLorenzo for suggesting improvements to the paper
and for helping us locate a subtle performance bug, and to the
Bellairs Research Institute of McGill University for providing a
wonderful research environment. This work was supported in part by
the National Science Foundation under grants NeTS-1413972 and
AiTF-1637532, by the European Research Council under grant~679127, by
a Facebook TAV award, by a Royal Society Wolfson fellowship, and a
gift from Keysight.
\end{acks}

\balance
\bibliography{pldi,bdds}

\ifextended
  \newpage
  \appendix
  \label{appendix}
  \onecolumn
  \include{appendix}
\fi
 
\end{document}

%% file: chain-sketch.tex
\begin{tikzpicture}[->,>=stealth,auto,semithick]
  \tikzstyle{config}=[node distance=2.2em]
  \tikzstyle{edge}=[]

  \node[config]   (C1)                  {\config{\polp\star, a, b}};
  \node[config]   (C2) [right = of C1]  {\config{\ptrue\pcomp\pseq{\polp}{\polp\star}, a, b}};
  \node[config]   (C3) [right = of C2]  {\config{\pseq{\polp}{\polp\star}, a, b \cup a}};
  \node[config]   (C4) [below = of C3, yshift=-1em]  {\config{\polp\star, a', b \cup a}};

  \path (C1) edge [dashed, thick]  node [edge] {$1$}                   (C2)
        (C2) edge [dashed, thick]  node [edge] {$1$}                   (C3)
        (C3) edge [dashed, thick]  node [edge] {$\bden{\polp}_{a,a'}$} (C4)
        (C1) edge [sloped]         node [edge, xshift=2em]
          {$\bden{\polp}_{a,a'}$} ([yshift=-1ex]C4.north west);
\end{tikzpicture}

%% file: img/pipeline.tex
\usetikzlibrary{backgrounds,calc,shapes.arrows,shapes.symbols,positioning,arrows.meta}
\definecolor{fddnodecol}{HTML}{111111}
\definecolor{fddleafcol}{HTML}{111111}

\tikzset{fddnode/.style={
    circle,draw=fddnodecol,fill=white,
    minimum width=0.2cm,
    minimum height=0.2cm,
  },
  fddleaf/.style={
    rectangle,draw=fddleafcol,fill=white,
    minimum width=0.2cm,
    minimum height=0.2cm,
  },
}

\begin{tikzpicture}[scale=0.9, every node/.style={scale=1.0}]
 \path[use as bounding box] (-2,-1.8) rectangle (10.5,3.5);

\begin{scope}[xshift=-1cm, yshift=1.5cm]
    \newcommand{\keyword}[1]{\textbf{\textcolor{blue}{#1}}}
    \node (code) [fill=gray!10,text width=2.7cm]{\small
        \texttt{\keyword{if} $\match{\pt}{1}$ \keyword{then} \\
              \quad $\modify{\pt}{2}$ $\oplus_{0.5}$ $\modify{\pt}{3}$ \\
            \keyword{else} \keyword{if} $\match{\pt}{2}$  \keyword{then}\\
              \quad $\modify{\pt}{1}$ \\
            \keyword{else} \keyword{if} $\match{\pt}{3}$ \keyword{then} \\
              \quad $\modify{\pt}{1}$ \\
            \keyword{else} \\
              \quad $\pfalse$
        }
    };
\end{scope}

\begin{scope}[xshift=2.5cm,scale=0.7, every node/.style={scale=0.7}]
 \node[fddleaf, label={below:$\modify{\pt}{2} \oplus_{0.5} \modify{\pt}{3}$}](leaf 1){};

 \node[fddleaf, right of=leaf 1, xshift=1.2cm,label={below:$\modify{\pt}{1}$}](leaf 2){};
 \node[fddleaf, right of=leaf 2, xshift=1cm,label={below:$\pfalse$}](leaf 3){};

 \node[fddnode, above of=leaf 1, xshift=3cm, yshift=0.5cm, label={[yshift=-0mm]left:$\match{\pt}{3}$}](node 3){};

 \node[fddnode, above of=leaf 1, xshift=1.5cm, yshift=2.0cm, label={[yshift=-2mm]left:$\match{\pt}{2}$}](node 2){};
 \node[fddnode, above of=leaf 1, xshift=0cm, yshift=3.5cm, label={left:$\match{\pt}{1}$}](node 1){};

 \draw[-latex,thick,black!50!gray] (node 1)--(leaf 1) node[]{} ;
 \draw[dashed,-latex,thick,black!50!gray] (node 1)--(node 2) node[]{} ;

 \draw[-latex,thick,black!50!gray] (node 2)--(leaf 2) node[]{} ;
 \draw[dashed,-latex,thick,black!50!gray] (node 2)--(node 3) node[]{} ;

 \draw[-latex,thick,black!50!gray] (node 3)--(leaf 2) node[]{} ;
 \draw[dashed,-latex,thick,black!50!gray] (node 3)--(leaf 3) node[]{} ;

\end{scope}

\begin{scope}[xshift=8cm, yshift=0.3cm]
 \node[yshift=1cm](matrix){
\setlength{\kbcolsep}{0pt}
\setlength{\kbrowsep}{0pt}
\setlength{\arraycolsep}{2pt}
\def\arraystretch{1.2}

\kbordermatrix{
              & \emptyset & \match{\pt}{1}  & \match{\pt}{2}  & \match{\pt}{3}  & \match{\pt}{*} \\
\emptyset     & 1         &                  &                &                 &         \\
\match{\pt}{1}&           &                  & \frac{1}{2}    & \frac{1}{2}     &         \\
\match{\pt}{2}&           & 1                &                &                 &         \\
\match{\pt}{3}&           & 1                &                &                 &         \\
\match{\pt}{*}& 1         &                  &                &                 &  
}};
\end{scope}

\node(code label)[below of=code,yshift=-1.5cm]{Program};
\node(fdd label)[right of=code label,xshift=3.5cm]{Probabilistic FDD};
\node(matrix label)[right of=fdd label,xshift=4.0cm]{Sparse matrix};

\draw [->,line width=0.5mm,>=stealth,shorten >=0.2cm,shorten <=0.2cm] ([xshift=2mm]code label.east)--([xshift=-2mm]fdd label.west) node[midway,above]
  {\small \textbf{Compile}};
\draw [-{>[left]},line width=0.5mm,>=stealth,shorten >=0.2cm,shorten <=0.2cm] ([xshift=2mm,yshift=1mm]fdd label.east)--([xshift=-2mm,yshift=1mm]matrix label.west) node[midway,above]{\small \textbf{Convert}};
\draw [-{>[left]},line width=0.5mm,>=stealth,shorten >=0.2cm,shorten <=0.2cm] ([xshift=-2mm,yshift=-1mm]matrix label.west)--([xshift=2mm,yshift=-1mm]fdd label.east);
\draw [-{>[left]},line width=0.5mm,>=stealth,shorten >=0.1cm,shorten <=0.1cm] (matrix label) to [out=-25,in=25,loop,looseness=6] node[right,above,xshift=-5.5mm,yshift=3.7mm] {\small \textbf{Solve}} (matrix label);


\end{tikzpicture}

%% file: img/fattree.tex
\usetikzlibrary{backgrounds,calc,shapes.arrows,shapes.symbols,positioning}
\definecolor{switchcol}{HTML}{999999}
\definecolor{switchacol}{HTML}{5EB3E5}
\definecolor{switchbcol}{HTML}{EE7355}
\makeatletter
\pgfkeys{/pgf/.cd,
  parallelepiped offset x/.initial=2mm,
  parallelepiped offset y/.initial=2mm
}
\pgfdeclareshape{parallelepiped}
{
  \inheritsavedanchors[from=rectangle] 
  \inheritanchorborder[from=rectangle]
  \inheritanchor[from=rectangle]{north}
  \inheritanchor[from=rectangle]{north west}
  \inheritanchor[from=rectangle]{north east}
  \inheritanchor[from=rectangle]{center}
  \inheritanchor[from=rectangle]{west}
  \inheritanchor[from=rectangle]{east}
  \inheritanchor[from=rectangle]{mid}
  \inheritanchor[from=rectangle]{mid west}
  \inheritanchor[from=rectangle]{mid east}
  \inheritanchor[from=rectangle]{base}
  \inheritanchor[from=rectangle]{base west}
  \inheritanchor[from=rectangle]{base east}
  \inheritanchor[from=rectangle]{south}
  \inheritanchor[from=rectangle]{south west}
  \inheritanchor[from=rectangle]{south east}
  \backgroundpath{
    \southwest \pgf@xa=\pgf@x \pgf@ya=\pgf@y
    \northeast \pgf@xb=\pgf@x \pgf@yb=\pgf@y
    \pgfmathsetlength\pgfutil@tempdima{\pgfkeysvalueof{/pgf/parallelepiped
      offset x}}
    \pgfmathsetlength\pgfutil@tempdimb{\pgfkeysvalueof{/pgf/parallelepiped
      offset y}}
    \def\ppd@offset{\pgfpoint{\pgfutil@tempdima}{\pgfutil@tempdimb}}
    \pgfpathmoveto{\pgfqpoint{\pgf@xa}{\pgf@ya}}
    \pgfpathlineto{\pgfqpoint{\pgf@xb}{\pgf@ya}}
    \pgfpathlineto{\pgfqpoint{\pgf@xb}{\pgf@yb}}
    \pgfpathlineto{\pgfqpoint{\pgf@xa}{\pgf@yb}}
    \pgfpathclose
    \pgfpathmoveto{\pgfqpoint{\pgf@xb}{\pgf@ya}}
    \pgfpathlineto{\pgfpointadd{\pgfpoint{\pgf@xb}{\pgf@ya}}{\ppd@offset}}
    \pgfpathlineto{\pgfpointadd{\pgfpoint{\pgf@xb}{\pgf@yb}}{\ppd@offset}}
    \pgfpathlineto{\pgfpointadd{\pgfpoint{\pgf@xa}{\pgf@yb}}{\ppd@offset}}
    \pgfpathlineto{\pgfqpoint{\pgf@xa}{\pgf@yb}}
    \pgfpathmoveto{\pgfqpoint{\pgf@xb}{\pgf@yb}}
    \pgfpathlineto{\pgfpointadd{\pgfpoint{\pgf@xb}{\pgf@yb}}{\ppd@offset}}
  }
}
\makeatother

\tikzset{switcha/.style={
    parallelepiped,fill=switchacol,draw=white,
    minimum width=0.75cm,
    minimum height=0.45cm,
    parallelepiped offset x=1.75mm,
    parallelepiped offset y=1.25mm,
    path picture={
      \draw[top color=switchacol!50,bottom color=switchacol!100]
      (path picture bounding box.south west) rectangle
      (path picture bounding box.north east);
      \draw[->,white,thick] ([xshift=3.75mm,yshift=-1.05mm]path picture bounding box.west) -- ([xshift=-2.75mm,yshift=-1.05mm]path picture bounding box.east);
      \draw[->,white,thick] ([xshift=-4.75mm,yshift=0.25mm]path picture bounding box.east) -- ([xshift=1.75mm,yshift=0.25mm]path picture bounding box.west);
     }
  },
  switchb/.style={
    parallelepiped,fill=switchbcol,draw=white,
    minimum width=0.75cm,
    minimum height=0.45cm,
    parallelepiped offset x=1.75mm,
    parallelepiped offset y=1.25mm,
    path picture={
      \draw[top color=switchbcol!50,bottom color=switchbcol!100]
      (path picture bounding box.south west) rectangle
      (path picture bounding box.north east);
      \draw[->,white,thick] ([xshift=3.75mm,yshift=-1.05mm]path picture bounding box.west) -- ([xshift=-2.75mm,yshift=-1.05mm]path picture bounding box.east);
      \draw[->,white,thick] ([xshift=-4.75mm,yshift=0.25mm]path picture bounding box.east) -- ([xshift=1.75mm,yshift=0.25mm]path picture bounding box.west);
     }
  },
  switch/.style={
    parallelepiped,fill=switchcol,draw=white,
    minimum width=0.75cm,
    minimum height=0.45cm,
    parallelepiped offset x=1.75mm,
    parallelepiped offset y=1.25mm,
    path picture={
      \draw[top color=switchcol!50,bottom color=switchcol!100]
      (path picture bounding box.south west) rectangle
      (path picture bounding box.north east);
      \draw[->,white,thick] ([xshift=3.75mm,yshift=-1.05mm]path picture bounding box.west) -- ([xshift=-2.75mm,yshift=-1.05mm]path picture bounding box.east);
      \draw[->,white,thick] ([xshift=-4.75mm,yshift=0.25mm]path picture bounding box.east) -- ([xshift=1.75mm,yshift=0.25mm]path picture bounding box.west);
     }
  },
  server/.style={
    parallelepiped,
    fill=white, draw,
    minimum width=0.35cm,
    minimum height=0.75cm,
    parallelepiped offset x=3mm,
    parallelepiped offset y=2mm,
    path picture={
      \draw[top color=gray!5,bottom color=gray!40]
      (path picture bounding box.south west) rectangle
      (path picture bounding box.north east);
      \node[rectangle,fill=black, minimum height=0.5mm, minimum width=2mm, inner sep=0pt, ] at ([xshift=-1.5mm,yshift=1.5mm]path picture bounding box.center){};
      \node[rectangle,fill=black, minimum height=0.5mm, minimum width=2mm, inner sep=0pt, ] at ([xshift=-1.5mm,yshift=0.5mm]path picture bounding box.center){};
    }
  },
}

\begin{tikzpicture}[scale=0.72, every node/.style={scale=0.72}]
 \path[use as bounding box] (-0.45,-0.6) rectangle (8.0,4.6);
\node[switcha,label={below:s1}](edge 1){};
\node[switcha,xshift=0.00cm, right of= edge 1,label={below:s2}](edge 2){};

\node[switcha,above of=edge 1,xshift=0.0cm,yshift=0.5cm,label={[label distance=2mm]right:}] (agg 1){};
\node[switcha,above of=edge 2,xshift=0.0cm,yshift=0.5cm,label={[label distance=2mm]right:}] (agg 2){};

\draw[thick,darkgray!10!gray] ([yshift=1mm]edge 1.north)--(agg 1.south) node[pos=0.8,left,color=gray]{} ;
\draw[thick,darkgray!10!gray] ([yshift=1mm]edge 2.north)--(agg 1.south) node[pos=0.8,right,color=gray]{};
\draw[thick,darkgray!10!gray] ([yshift=1mm]edge 1.north)--(agg 2.south) node[pos=0.8,left,color=gray]{} ;
\draw[thick,darkgray!10!gray] ([yshift=1mm]edge 2.north)--(agg 2.south) node[pos=0.8,right,color=gray]{};

\begin{scope}[xshift=2cm]
  \node[switcha,label={below:s3}](edge 3){};
  \node[switcha,xshift=0.00cm, right of=edge 3,label={below:s4}](edge 4){};

  \node[switcha, above of=edge 3,xshift=0.0cm,yshift=0.5cm,label={[label distance=2mm]right:}] (agg 3){};
  \node[switcha, above of=edge 4,xshift=0.0cm,yshift=0.5cm,label={[label distance=2mm]right:}] (agg 4){};

  \draw[thick,darkgray!10!gray] ([yshift=1mm]edge 3.north)--(agg 3.south) node[pos=0.8,left,color=gray]{} ;
  \draw[thick,darkgray!10!gray] ([yshift=1mm]edge 4.north)--(agg 3.south) node[pos=0.8,right,color=gray]{};
  \draw[thick,darkgray!10!gray] ([yshift=1mm]edge 3.north)--(agg 4.south) node[pos=0.8,left,color=gray]{} ;
  \draw[thick,darkgray!10!gray] ([yshift=1mm]edge 4.north)--(agg 4.south) node[pos=0.8,right,color=gray]{};

\end{scope}

\begin{scope}[xshift=4cm]
  \node[switcha,label={below:s5}](edge 5){};
  \node[switcha,xshift=0.00cm, right of= edge 5,label={below:s6}](edge 6){};

  \node[switcha, above of=edge 5,xshift=0.0cm,yshift=0.5cm,label={[label distance=2mm]right:}] (agg 5){};
  \node[switcha, above of=edge 6,xshift=0.0cm,yshift=0.5cm,label={[label distance=2mm]right:}] (agg 6){};

  \draw[thick,darkgray!10!gray] ([yshift=1mm]edge 5.north)--(agg 5.south) node[pos=0.8,left,color=gray]{} ;
  \draw[thick,darkgray!10!gray] ([yshift=1mm]edge 6.north)--(agg 5.south) node[pos=0.8,right,color=gray]{};
  \draw[thick,darkgray!10!gray] ([yshift=1mm]edge 5.north)--(agg 6.south) node[pos=0.8,left,color=gray]{} ;
  \draw[thick,darkgray!10!gray] ([yshift=1mm]edge 6.north)--(agg 6.south) node[pos=0.8,right,color=gray]{};

\end{scope}

\begin{scope}[xshift=6cm]
  \node[switcha,label={below:s7}](edge 7){};
  \node[switcha,xshift=0.00cm, right of= edge 7,label={below:s8}](edge 8){};

  \node[switcha, above of=edge 7,xshift=0.0cm,yshift=0.5cm,label={[label distance=2mm]right:}] (agg 7){};
  \node[switcha, above of=edge 8,xshift=0.0cm,yshift=0.5cm,label={[label distance=2mm]right:}] (agg 8){};

  \draw[thick,darkgray!10!gray] ([yshift=1mm]edge 7.north)--(agg 7.south) node[pos=0.8,left,color=gray]{} ;
  \draw[thick,darkgray!10!gray] ([yshift=1mm]edge 8.north)--(agg 7.south) node[pos=0.8,right,color=gray]{} ;
  \draw[thick,darkgray!10!gray] ([yshift=1mm]edge 7.north)--(agg 8.south) node[pos=0.8,left,color=gray]{} ;
  \draw[thick,darkgray!10!gray] ([yshift=1mm]edge 8.north)--(agg 8.south) node[pos=0.8,right,color=gray]{} ;

\end{scope}

\node[switch, above of=agg 1, xshift=0.5cm,yshift=1.5cm,label={[label distance=2mm]right:}] (core 1){};
\node[switch, above of=agg 3, xshift=0.5cm,yshift=1.5cm,label={[label distance=2mm]right:}] (core 2){};
\node[switch, above of=agg 5, xshift=0.5cm,yshift=1.5cm,label={[label distance=2mm]right:}] (core 3){};
\node[switch, above of=agg 7, xshift=0.5cm,yshift=1.5cm,label={[label distance=2mm]right:}] (core 4){};



\draw[thick,darkgray!10!gray] ([yshift=1mm]agg 1.north)--(core 1.south) node[pos=0.9,left,color=gray]{} node[pos=0.0,above,color=gray]{};
\draw[thick,darkgray!10!gray] ([yshift=1mm]agg 1.north)--(core 2.south) node[pos=0.9,right,color=gray]{} node[pos=0.0,above,color=gray]{};

\draw[thick,darkgray!10!gray] ([yshift=1mm]agg 2.north)--(core 3.south) node[pos=0.9,left,color=gray]{} node[pos=0.0,above,color=gray]{};
\draw[thick,darkgray!10!gray] ([yshift=1mm]agg 2.north)--(core 4.south) node[pos=0.9,right,color=gray]{} node[pos=0.0,above,color=gray]{};

\draw[thick,darkgray!10!gray] ([yshift=1mm]agg 3.north)--(core 1.south) node[pos=0.9,left,color=gray]{} node[pos=0.0,above,color=gray]{};
\draw[thick,darkgray!10!gray] ([yshift=1mm]agg 3.north)--(core 2.south) node[pos=0.9,right,color=gray]{} node[pos=0.0,above,color=gray]{};

\draw[thick,darkgray!10!gray] ([yshift=1mm]agg 4.north)--(core 3.south) node[pos=0.9,left,color=gray]{} node[pos=0.0,above,color=gray]{};
\draw[thick,darkgray!10!gray] ([yshift=1mm]agg 4.north)--(core 4.south) node[pos=0.9,right,color=gray]{} node[pos=0.0,above,color=gray]{};

\draw[thick,darkgray!10!gray] ([yshift=1mm]agg 5.north)--(core 1.south) node[pos=0.9,left,color=gray]{} node[pos=0.0,above,color=gray]{};
\draw[thick,darkgray!10!gray] ([yshift=1mm]agg 5.north)--(core 2.south) node[pos=0.9,right,color=gray]{} node[pos=0.0,above,color=gray]{};

\draw[thick,darkgray!10!gray] ([yshift=1mm]agg 6.north)--(core 3.south) node[pos=0.9,left,color=gray]{} node[pos=0.0,above,color=gray]{};
\draw[thick,darkgray!10!gray] ([yshift=1mm]agg 6.north)--(core 4.south) node[pos=0.9,right,color=gray]{} node[pos=0.0,above,color=gray]{};

\draw[thick,darkgray!10!gray] ([yshift=1mm]agg 7.north)--(core 1.south) node[pos=0.9,left,color=gray]{} node[pos=0.0,above,color=gray]{};
\draw[thick,darkgray!10!gray] ([yshift=1mm]agg 7.north)--(core 2.south) node[pos=0.9,right,color=gray]{} node[pos=0.0,above,color=gray]{};

\draw[thick,darkgray!10!gray] ([yshift=1mm]agg 8.north)--(core 3.south) node[pos=0.9,left,color=gray]{} node[pos=0.0,above,color=gray]{};
\draw[thick,darkgray!10!gray] ([yshift=1mm]agg 8.north)--(core 4.south) node[pos=0.9,right,color=gray]{} node[pos=0.0,above,color=gray]{};
s

\end{tikzpicture}

%% file: img/abfattree.tex
\usetikzlibrary{backgrounds,calc,shapes.arrows,shapes.symbols,positioning}
\definecolor{switchcol}{HTML}{999999}
\definecolor{switchacol}{HTML}{5EB3E5}
\definecolor{switchbcol}{HTML}{EE7355}

\makeatletter
\pgfkeys{/pgf/.cd,
  parallelepiped offset x/.initial=2mm,
  parallelepiped offset y/.initial=2mm
}
\pgfdeclareshape{parallelepiped}
{
  \inheritsavedanchors[from=rectangle] 
  \inheritanchorborder[from=rectangle]
  \inheritanchor[from=rectangle]{north}
  \inheritanchor[from=rectangle]{north west}
  \inheritanchor[from=rectangle]{north east}
  \inheritanchor[from=rectangle]{center}
  \inheritanchor[from=rectangle]{west}
  \inheritanchor[from=rectangle]{east}
  \inheritanchor[from=rectangle]{mid}
  \inheritanchor[from=rectangle]{mid west}
  \inheritanchor[from=rectangle]{mid east}
  \inheritanchor[from=rectangle]{base}
  \inheritanchor[from=rectangle]{base west}
  \inheritanchor[from=rectangle]{base east}
  \inheritanchor[from=rectangle]{south}
  \inheritanchor[from=rectangle]{south west}
  \inheritanchor[from=rectangle]{south east}
  \backgroundpath{
    \southwest \pgf@xa=\pgf@x \pgf@ya=\pgf@y
    \northeast \pgf@xb=\pgf@x \pgf@yb=\pgf@y
    \pgfmathsetlength\pgfutil@tempdima{\pgfkeysvalueof{/pgf/parallelepiped
      offset x}}
    \pgfmathsetlength\pgfutil@tempdimb{\pgfkeysvalueof{/pgf/parallelepiped
      offset y}}
    \def\ppd@offset{\pgfpoint{\pgfutil@tempdima}{\pgfutil@tempdimb}}
    \pgfpathmoveto{\pgfqpoint{\pgf@xa}{\pgf@ya}}
    \pgfpathlineto{\pgfqpoint{\pgf@xb}{\pgf@ya}}
    \pgfpathlineto{\pgfqpoint{\pgf@xb}{\pgf@yb}}
    \pgfpathlineto{\pgfqpoint{\pgf@xa}{\pgf@yb}}
    \pgfpathclose
    \pgfpathmoveto{\pgfqpoint{\pgf@xb}{\pgf@ya}}
    \pgfpathlineto{\pgfpointadd{\pgfpoint{\pgf@xb}{\pgf@ya}}{\ppd@offset}}
    \pgfpathlineto{\pgfpointadd{\pgfpoint{\pgf@xb}{\pgf@yb}}{\ppd@offset}}
    \pgfpathlineto{\pgfpointadd{\pgfpoint{\pgf@xa}{\pgf@yb}}{\ppd@offset}}
    \pgfpathlineto{\pgfqpoint{\pgf@xa}{\pgf@yb}}
    \pgfpathmoveto{\pgfqpoint{\pgf@xb}{\pgf@yb}}
    \pgfpathlineto{\pgfpointadd{\pgfpoint{\pgf@xb}{\pgf@yb}}{\ppd@offset}}
  }
}
\makeatother

\tikzset{switcha/.style={
    parallelepiped,fill=switchacol,draw=white,
    minimum width=0.75cm,
    minimum height=0.45cm,
    parallelepiped offset x=1.75mm,
    parallelepiped offset y=1.25mm,
    path picture={
      \draw[top color=switchacol!50,bottom color=switchacol!100]
      (path picture bounding box.south west) rectangle
      (path picture bounding box.north east);
      \draw[->,white,thick] ([xshift=3.75mm,yshift=-1.05mm]path picture bounding box.west) -- ([xshift=-2.75mm,yshift=-1.05mm]path picture bounding box.east);
      \draw[->,white,thick] ([xshift=-4.75mm,yshift=0.25mm]path picture bounding box.east) -- ([xshift=1.75mm,yshift=0.25mm]path picture bounding box.west);
     }
  },
  switchb/.style={
    parallelepiped,fill=switchbcol,draw=white,
    minimum width=0.75cm,
    minimum height=0.45cm,
    parallelepiped offset x=1.75mm,
    parallelepiped offset y=1.25mm,
    path picture={
      \draw[top color=switchbcol!50,bottom color=switchbcol!100]
      (path picture bounding box.south west) rectangle
      (path picture bounding box.north east);
      \draw[->,white,thick] ([xshift=3.75mm,yshift=-1.05mm]path picture bounding box.west) -- ([xshift=-2.75mm,yshift=-1.05mm]path picture bounding box.east);
      \draw[->,white,thick] ([xshift=-4.75mm,yshift=0.25mm]path picture bounding box.east) -- ([xshift=1.75mm,yshift=0.25mm]path picture bounding box.west);
     }
  },
  switch/.style={
    parallelepiped,fill=switchcol,draw=white,
    minimum width=0.75cm,
    minimum height=0.45cm,
    parallelepiped offset x=1.75mm,
    parallelepiped offset y=1.25mm,
    path picture={
      \draw[top color=switchcol!50,bottom color=switchcol!100]
      (path picture bounding box.south west) rectangle
      (path picture bounding box.north east);
      \draw[->,white,thick] ([xshift=3.75mm,yshift=-1.05mm]path picture bounding box.west) -- ([xshift=-2.75mm,yshift=-1.05mm]path picture bounding box.east);
      \draw[->,white,thick] ([xshift=-4.75mm,yshift=0.25mm]path picture bounding box.east) -- ([xshift=1.75mm,yshift=0.25mm]path picture bounding box.west);
     }
  },
  server/.style={
    parallelepiped,
    fill=white, draw,
    minimum width=0.35cm,
    minimum height=0.75cm,
    parallelepiped offset x=3mm,
    parallelepiped offset y=2mm,
    path picture={
      \draw[top color=gray!5,bottom color=gray!40]
      (path picture bounding box.south west) rectangle
      (path picture bounding box.north east);
      \node[rectangle,fill=black, minimum height=0.5mm, minimum width=2mm, inner sep=0pt, ] at ([xshift=-1.5mm,yshift=1.5mm]path picture bounding box.center){};
      \node[rectangle,fill=black, minimum height=0.5mm, minimum width=2mm, inner sep=0pt, ] at ([xshift=-1.5mm,yshift=0.5mm]path picture bounding box.center){};
    }
  },
}

\begin{tikzpicture}[scale=0.8, every node/.style={scale=0.8}]

\node[switcha,label={below:s1}](edge 1){};
\node[switcha,xshift=0.00cm, right of= edge 1,label={below:s2}](edge 2){};

\node[switcha,above of=edge 1,xshift=0.0cm,yshift=0.5cm,label={[label distance=0mm, xshift=-0.3cm]below:$A$}] (agg 1){};
\node[switcha,above of=edge 2,xshift=0.0cm,yshift=0.5cm,label={[label distance=2mm]right:}] (agg 2){};

\draw[latex-,ultra thick,black] ([yshift=1mm]edge 1.north)--(agg 1.south) node[pos=0.8,left,color=gray]{} ;
\draw[thin,gray] ([yshift=1mm]edge 2.north)--(agg 1.south) node[pos=0.8,right,color=gray]{};
\draw[latex-,thick,blue!80!black] ([xshift=-1mm,yshift=1mm]edge 1.north)--([xshift=-1mm]agg 2.south) node[pos=0.8,left,color=gray]{} ;
\draw[latex-,thick,red!90!gray] ([yshift=1mm]edge 1.north)--(agg 2.south) node[pos=0.8,left,color=gray]{} ;
\draw[thin,gray] ([yshift=1mm]edge 2.north)--(agg 2.south) node[pos=0.8,right,color=gray]{};

\begin{scope}[xshift=2cm]
  \node[switchb,label={below:s3}](edge 3){};
  \node[switchb,xshift=0.00cm, right of=edge 3,label={below:s4}](edge 4){};

  \node[switchb, above of=edge 3,xshift=0.0cm,yshift=0.5cm,label={[label distance=0mm,xshift=-0.3cm]below:$A'$}] (agg 3){};
  \node[switchb, above of=edge 4,xshift=0.0cm,yshift=0.5cm,label={[label distance=2mm]right:}] (agg 4){};

  \draw[thin,gray] ([yshift=1mm]edge 3.north)--(agg 3.south) node[pos=0.8,left,color=gray]{} ;
  \draw[thin,gray] ([yshift=1mm]edge 4.north)--(agg 3.south) node[pos=0.8,right,color=gray]{};
  \draw[thin,gray] ([yshift=1mm]edge 3.north)--(agg 4.south) node[pos=0.8,left,color=gray]{} ;
  \draw[thin,gray] ([yshift=1mm]edge 4.north)--(agg 4.south) node[pos=0.8,right,color=gray]{};

\end{scope}

\begin{scope}[xshift=4cm]
  \node[switcha,label={below:s5}](edge 5){};
  \node[switcha,xshift=0.00cm, right of= edge 5,label={below:s6}](edge 6){};

  \node[switcha, above of=edge 5,xshift=0.0cm,yshift=0.5cm,label={[label distance=0mm, xshift=-0.3cm]below:$A''$}] (agg 5){};
  \node[switcha, above of=edge 6,xshift=0.0cm,yshift=0.5cm,label={[label distance=2mm]right:}] (agg 6){};

  \draw[latex-,thick,red!90!gray] ([yshift=1mm]edge 5.north)--(agg 5.south) node[pos=0.8,left,color=gray]{} ;
  \draw[thin,gray] ([yshift=1mm]edge 6.north)--(agg 5.south) node[pos=0.8,right,color=gray]{};
  \draw[-latex,thick,red!90!gray] ([yshift=1mm]edge 5.north)--(agg 6.south) node[pos=0.8,left,color=gray]{} ;
  \draw[thin,gray] ([yshift=1mm]edge 6.north)--(agg 6.south) node[pos=0.8,right,color=gray]{};

\end{scope}

\begin{scope}[xshift=6cm]
  \node[switchb,label={below:s7}](edge 7){};
  \node[switchb,xshift=0.00cm, right of= edge 7,label={below:s8}](edge 8){};

  \node[switchb, above of=edge 7,xshift=0.0cm,yshift=0.5cm,label={[label distance=2mm]right:}] (agg 7){};
  \node[switchb, above of=edge 8,xshift=0.0cm,yshift=0.5cm,label={[label distance=2mm]right:}] (agg 8){};

  \draw[-latex,ultra thick,black] ([yshift=1mm]edge 7.north)--(agg 7.south) node[pos=0.8,left,color=gray]{} ;
  \draw[thin,gray] ([yshift=1mm]edge 8.north)--(agg 7.south) node[pos=0.8,right,color=gray]{} ;
  \draw[thin,gray] ([yshift=1mm]edge 7.north)--(agg 8.south) node[pos=0.8,left,color=gray]{} ;
  \draw[thin,gray] ([yshift=1mm]edge 8.north)--(agg 8.south) node[pos=0.8,right,color=gray]{} ;

\end{scope}

\node[switch, above of=agg 1, xshift=0.5cm,yshift=1.5cm,label={[label distance=0mm, xshift=-0.3cm]below:$C$}] (core 1){};
\node[switch, above of=agg 3, xshift=0.5cm,yshift=1.5cm,label={[label distance=2mm]right:}] (core 2){};
\node[switch, above of=agg 5, xshift=0.5cm,yshift=1.5cm,label={[label distance=2mm]right:}] (core 3){};
\node[switch, above of=agg 7, xshift=0.5cm,yshift=1.5cm,label={[label distance=2mm]right:}] (core 4){};


\node[xshift=1.0cm,yshift=0.0cm,left of = agg 1,align=center,color=red](lev1) {\Huge \xmark};

\draw[latex-,ultra thick,black] ([yshift=1mm]agg 1.north)--(core 1.south) node[pos=0.9,left,color=gray]{} node[pos=0.0,above,color=gray]{};
\draw[thin,gray] ([yshift=1mm]agg 1.north)--(core 2.south) node[pos=0.9,right,color=gray]{} node[pos=0.0,above,color=gray]{};

\draw[latex-,thick,blue!80!black] ([yshift=1mm,xshift=1mm]agg 2.north)--(core 3.south) node[pos=0.9,left,color=gray]{} node[pos=0.0,above,color=gray]{};
\draw[latex-,thick,red!90!gray] ([yshift=1mm,xshift=2.5mm]agg 2.north)--(core 4.south) node[pos=0.9,right,color=gray]{} node[pos=0.0,above,color=gray]{};

\draw[latex-,thick,blue!80!black] ([yshift=1mm]agg 3.north)--(core 1.south) node[pos=0.9,left,color=gray]{} node[pos=0.0,above,color=gray]{};
\draw[-latex,thick,blue!80!black] ([yshift=1mm]agg 3.north)--(core 3.south) node[pos=0.9,right,color=gray]{} node[pos=0.0,above,color=gray]{};

\draw[thin,gray] ([yshift=1mm]agg 4.north)--(core 2.south) node[pos=0.9,left,color=gray]{} node[pos=0.0,above,color=gray]{};
\draw[thin,gray] ([yshift=1mm]agg 4.north)--(core 4.south) node[pos=0.9,right,color=gray]{} node[pos=0.0,above,color=gray]{};

\draw[latex-,thick,red!90!gray] ([yshift=1mm]agg 5.north)--(core 1.south) node[pos=0.9,left,color=gray]{} node[pos=0.0,above,color=gray]{};
\draw[thin,gray] ([yshift=1mm]agg 5.north)--(core 2.south) node[pos=0.9,right,color=gray]{} node[pos=0.0,above,color=gray]{};

\draw[thin,gray] ([yshift=1mm]agg 6.north)--(core 3.south) node[pos=0.9,left,color=gray]{} node[pos=0.0,above,color=gray]{};
\draw[-latex,thick,red!90!gray] ([yshift=1mm]agg 6.north)--(core 4.south) node[pos=0.9,right,color=gray]{} node[pos=0.0,above,color=gray]{};

\draw[-latex,ultra thick,black] ([yshift=1mm]agg 7.north)--(core 1.south) node[pos=0.9,left,color=gray]{} node[pos=0.0,above,color=gray]{};
\draw[thin,gray] ([yshift=1mm]agg 7.north)--(core 3.south) node[pos=0.9,right,color=gray]{} node[pos=0.0,above,color=gray]{};

\draw[thin,gray] ([yshift=1mm]agg 8.north)--(core 2.south) node[pos=0.9,left,color=gray]{} node[pos=0.0,above,color=gray]{};
\draw[thin,gray] ([yshift=1mm]agg 8.north)--(core 4.south) node[pos=0.9,right,color=gray]{} node[pos=0.0,above,color=gray]{};

\end{tikzpicture}

%% file: appendix.tex
\section{\probnetkat Denotational Semantics}
\label{app:semantics}

\begin{figure}[t!]
\begin{tabular}{c|c}
\textbf{Semantics}\quad\fbox{\(\den{\polp} \in \pset{\Pk} \to \Dist(\pset{\Pk})\)}&
\textbf{(Discrete) Probability Monad} \(\Dist\)\\
\def\arraystretch{1.25}
$\begin{array}{r@{~~}c@{~~}l}
\den{\pfalse}(a) & \defeq &
  \delta({\emptyset})\\
\den{\ptrue}(a) & \defeq &
  \delta({a})\\
\den{\match{\field}{n}}(a) & \defeq &
  \delta({\set{\pk \in a}{\pk.f = n}}) \\
\den{\modify{\field}{n}}(a) & \defeq &
  \delta({\set{\upd{\pk}{\field}{n}}{\pk \in a }}) \\
\den{\pnot{\preda}}(a) & \defeq &
  \Dist (\lambda b. a-b)(\den{\preda}(a))\\
\den{\punion{\polp}{\polq}}(a) & \defeq &
  \Dist(\cup)(\den{\polp}(a) \times \den{\polq}(a))\\
\den{\pseq{\polp}{\polq}}(a) & \defeq &
  \lift{\den{\polq}}(\den{\polp}(a))\\
\den{\polp \mathop{\opr} \polq}(a) & \defeq &
  r \cdot \den{\polp}(a) + (1-r) \cdot \den{\polq}(a)\\
\den{\pstar\polp}(a) & \defeq & \displaystyle\bigsqcup_{n \in \N} \den{p^{(n)}}(a)\\
\multicolumn{3}{l}{\text{where } ~ p^{(0)} \defeq \ptrue, \hspace{1.5ex}
  p^{(n+1)} \defeq \punion{\ptrue}{\pseq{p}{p^{(n)}}}}
\end{array}$
%
%

&\def\arraystretch{1.25}
$\begin{array}{@{}r@{\quad}l@{}}
\text{Unit} & \delta : X \to \Dist(X)\quad\delta(x) \defeq \dirac x\\
\text{Bind} & \lift{-} : (X \to \Dist(Y)) \to \Dist(X) \to \Dist(Y)\\
& \lift{f}(\mu)(A) \defeq \sum_{x \in X} f(x)(A) \cdot \mu(x)
\end{array}$
\end{tabular}
\vspace*{-1em}
\caption{\probnetkat semantics.}
\label{fig:probnetkat-app}
\end{figure}

In the original \probnetkat language, programs manipulate sets
of \emph{packet histories}---non-empty, finite sequences of packets
modeling trajectories through the
network~\citep{probnetkat-scott,probnetkat-cantor}. The resulting state
space is uncountable and modeling the semantics properly requires
full-blown measure theory as some programs generate continuous
distributions. In the history-free fragment, programs manipulate sets
of packets and the state space is finite, which makes the semantics
considerably simpler.

\begin{proposition}
\label{prop:old-and-new-sem}
Let $\oldden{-}$ denote the semantics defined
in \citet{probnetkat-scott}.  Then for all $\pdup$-free programs
$\polp$ and inputs $a \in \pPk$, we have
\(
  \den{\polp}(a) = \oldden{\polp}(a)
\), 
where we identify packets and histories of length one.
\end{proposition}

Throughout this paper, we can work in the discrete space $\pPk$, \ie,
the set of sets of packets. An \emph{outcome} (denoted by lowercase
variables $a,b,c,\dots$) is a set of packets and an \emph{event}
(denoted by uppercase variables $A,B,C,\dots$) is a set of
outcomes. Given a discrete probability measure on this space, the
probability of an event is the sum of the probabilities of its
outcomes.

\probnetkat programs are interpreted as \emph{Markov kernels} on the
space $\pPk$.  A Markov kernel is a function $\pPk \to \Dist(\pPk)$
where $\Dist$ is the probability (or Giry)
monad \citep{giry1982categorical,K81c}. Thus, a program $\polp$ maps an
input set of packets $a \in \pPk$ to a
\emph{distribution} $\den{\polp}(a) \in \Dist(\pPk)$ over output sets of packets.
The semantics uses the following probabilistic
constructions:\footnote{These can also be defined for uncountable
spaces, as would be required to handle the full language.}
\begin{itemize}
  \item For a discrete measurable space $X$, $\Dist(X)$ denotes the
  set of probability measures over $X$; that is, the set of countably
  additive functions $\mu : 2^X \to [0,1]$ with $\mu(X)=1$.
  \item For a measurable function $f : X \to Y$, $\Dist(f)
  : \Dist(X) \to \Dist(Y)$ denotes the \emph{pushforward} along $f$;
  that is, the function that maps a measure $\mu$ on $X$
  to \[ \Dist(f)(\mu) \defeq \mu \circ f^{-1} = \lambda
  A \in \Sigma_Y.\ \mu(\set{x \in X}{f(x) \in A}) \] which is called
  the \emph{pushforward measure} on $Y$.
  \item The \emph{unit} $\delta : X \to \Dist(X)$ of the monad maps a
  point $x \in X$ to the point mass (or \emph{Dirac} measure) $\dirac
  x \in \Dist(X)$.  The Dirac measure is given by \[ \dirac x
  (A) \defeq \ind{x \in A} \] That is, the Dirac measure is $1$ if
  $x \in A$ and $0$ otherwise.
  \item The \emph{bind} operation of the monad,
  \[
  \bind - : (X \to \Dist(Y)) \to \Dist(X) \to \Dist(Y)
  \]
  lifts a function $f: X \to \Dist(Y)$ with deterministic inputs to a
  function $\bind f: \Dist(X) \to \Dist(Y)$ that takes random
  inputs. Intuitively, this is achieved by averaging the output of $f$
  when the inputs are randomly distributed according to
  $\mu$. Formally,
  \begin{equation*}
    \bind f(\mu)(A) \defeq 
    \sum_{x \in X} f(x)(A) \cdot \mu(x).
  \end{equation*}
  \item Given two measures $\mu \in \Dist(X)$ and $\nu \in \Dist(Y)$,  
  $\mu\times\nu \in \Dist(X \times Y)$ denotes their \emph{product measure}.
  This is the unique measure satisfying
  \[
  (\mu\times\nu)(A \times B) = \mu(A) \cdot \nu(B)
  \]
  Intuitively, it models distributions over pairs of independent
  values.
\end{itemize}

Using these primitives, we can now make our operational intuitions
precise (see \cref{fig:probnetkat-app} for formal definitions). A
predicate $\preda$ maps the set of input packets $a \in \pPk$ to the
subset of packets $b \subseteq a$ satisfying the predicate (with
probability $1$). Hence, $\pfalse$ drops all packets (i.e., it returns
the empty set) while $\ptrue$ keeps all packets (i.e., it returns the
input set). The test $\match{\field}{n}$ returns the subset of input
packets whose $\field$-field is $n$.  Negation $\pnot{\preda}$ filters
out the packets returned by $\preda$.

Parallel composition $\punion{\polp}{\polq}$ executes $\polp$ and
$\polq$ independently on the input set, then returns the union of
their results. Note that packet sets do \emph{not} model
nondeterminism, unlike the usual situation in Kleene
algebras---rather, they model collections of packets traversing
possibly different portions of the network simultaneously.  In
particular, the union operation is \emph{not} idempotent:
$\punion{\polp}{\polp}$ need not have the same semantics as $\polp$.
Probabilistic choice $\polp \opr \polq$ feeds the input to both
$\polp$ and $\polq$ and returns a convex combination of the output
distributions according to $r$. Sequential composition
$\pseq{\polp}{\polq}$ can be thought of as a two-stage probabilistic
process: it first executes $\polp$ on the input set to obtain a random
intermediate result, then feeds that into $\polq$ to obtain the final
distribution over outputs. The outcome of $\polq$ is averaged over the
distribution of intermediate results produced by $\polp$.

We say that two programs are \emph{equivalent}, denoted
$\polp \equiv \polq$, if they denote the same Markov kernel, \ie if
$\den{\polp} = \den{\polq}$. As usual, we expect Kleene star
$\polp\star$ to satisfy the characteristic fixed point equation
$\polp\star \equiv \punion{\ptrue}{\pseq{\polp}{\polp\star}}$, which
allows it to be unrolled ad infinitum. Thus we define it as the
supremum of its finite unrollings $\polp^{(n)}$; see
\cref{fig:probnetkat-app}. This supremum is taken
in a CPO $(\Dist(\pPk), \sqleq)$ of distributions that is described in
more detail in \cref{sec:measurable-space}.  The partial ordering
$\sqleq$ on packet set distributions gives rise to a partial ordering
on programs: we write $\polp \leq \polq$ iff
$\den{\polp}(a) \sqleq \den{\polq}(a)$ for all inputs
$a \in \pPk$. Intuitively, $\polp \leq \polq$ iff $\polp$ produces any
particular output packet $\pi$ with probability at most that of
$\polq$ for any fixed input---$\polq$ has a larger probability of delivering
more output packets.

\subsection{The CPO \texorpdfstring{$(\Dist(\pPk), \sqleq)$}{of distributions}}
\label{sec:measurable-space}
The space $\pPk$ with the subset order forms a CPO
$(\pPk, \subseteq)$. Following \citet{Saheb-Djahromi80},
this CPO can be lifted to a CPO $(\Dist(\pPk), \sqleq)$ on
distributions over $\pPk$. Because $\pPk$ is a finite space, the
resulting ordering $\sqleq$ on distributions takes a particularly easy
form:
\[
\mu \sqleq \nu \quad \iff \quad \mu(\sset{a}{\uparrow}) \leq \nu(\sset{a}{\uparrow}) \text{ for all } a \subseteq \Pk
\]
where $\sset{a}{\uparrow} \defeq \set{b}{a \subseteq b}$ denotes
upward closure.  Intuitively, $\nu$ produces more outputs then $\mu$.
As was shown in \citet{probnetkat-scott}, \probnetkat satisfies various
monotonicity (and continuity) properties with respect to this
ordering, including
\begin{align*}
a \subseteq a' \ \implies \ \den{\polp}(a) \sqleq \den{\polp}(a')
\qquad \text{and} \qquad
n \leq m\ \implies \ \den{\polp^{(n)}}(a) \sqleq \den{\polp^{(m)}}(a).
\end{align*}
As a result, the semantics of $\polp\star$ as the supremum of its
finite unrollings $\polp^{(n)}$ is well-defined.

While the semantics of full \probnetkat requires more domain theory to give a
satisfactory characterization of Kleene star, a simpler characterization
suffices for the history-free fragment.
\begin{lemma}[Pointwise Convergence]
\label{lem:ptwise-convergence}
Let $A \subseteq \pPk$. Then for all programs $\polp$ and inputs
$a \in \pPk$,
\begin{align*}
  \den{\polp\star}(a)(A) = \lim_{n\to\infty} \den{\polp^{(n)}}(a)(A).
\end{align*}
\end{lemma}
%

\section{Omitted Proofs}
\label{app:omitted-proofs}

\begin{lemma}
\label{lem:convergence-on-cantor-open}
Let $A$ be a finite boolean combination of basic open sets, \ie sets of the form
$B_a = \sset{a}\uparrow$ for $a \in \pfin\Hist$, and let $\oldden{-}$ denote
the semantics from \citet{probnetkat-scott}. Then for all programs
$\polp$ and inputs $a \in \pH$,
\begin{align*}
  \oldden{\polp\star}(a)(A) = \lim_{n\to\infty} \oldden{\polp^{(n)}}(a)(A)
\end{align*}
\end{lemma}
\begin{proof*}
Using topological arguments, the claim follows directly from previous results:
$A$ is a Cantor-clopen set by \citet{probnetkat-scott} (\ie, both $A$ and
$\scomp{A}$ are Cantor-open), so its indicator function
$\mathbf{1}_A$ is Cantor-continuous. But $\mu_n \defeq \oldden{\polp^{(n)}}(a)$
converges weakly to $\mu \defeq \oldden{\polp\star}(a)$ in the Cantor topology
\citep[Theorem 4]{probnetkat-cantor}, so
\begin{align*}
\lim_{n\to\infty} \oldden{\polp^{(n)}}(a)(A)
= \lim_{n\to\infty} \int \mathbf{1}_A d\mu_n
= \int \mathbf{1}_A d\mu
= \oldden{\polp\star}(a)(A)
\end{align*}
(To see why $A$ and $\scomp{A}$ are open in the Cantor topology, note that they
can be written in disjunctive normal form over atoms $B_{\sset{h}}$.)
\end{proof*}

Predicates in \probnetkat form a Boolean algebra.

\begin{lemma}
\label{lem:preds-boolean-algebra}
Every predicate $t$ satisfies $\den{t}(a) = \dirac{a \cap b_t}$ for a
certain packet set
$b_t \subseteq \Pk$, where
\begin{itemize}
\item $b_\pfalse = \emptyset$,
\item $b_\ptrue = \Pk$,
\item $b_{\match{\field}{n}} = \set{\pk\in\Pk}{\pk.f=n}$,
\item $b_{\pnot{\preda}} = \Pk - b_{\preda}$,
\item $b_{\punions{\preda}{\predb}} = b_{\preda} \cup b_{\predb}$, and
\item $b_{\pseq{\preda}{\predb}} = b_{\preda} \cap b_{\predb}$.
\end{itemize}
\end{lemma}
\begin{proof*}
For $\pfalse$, $\ptrue$, and $\match{\field}{n}$, the claim holds trivially.
For $\pnot{\preda}$, $\punion{\preda}{\predb}$, and $\pseq{\preda}{\predb}$,
the claim follows inductively, using that
$\Dist(f)(\dirac b) = \dirac{f(b)}$,
$\dirac{b} \times \dirac{c} = \dirac{(b,c)}$, and that
$f^\dagger(\dirac{b}) = f(b)$. The first and last equations hold because
$\langle\Dist, \delta, \lift{-}\rangle$ is a monad.
\end{proof*}

\begin{proof}[Proof of \cref{prop:old-and-new-sem}]
We only need to show that for $\pdup$-free programs $\polp$ and history-free
inputs $a \in \pPk$, $\oldden{\polp}(a)$ is a distribution on packets (where 
we identify packets and singleton histories).
We proceed by structural induction on $\polp$. All cases are straightforward
except perhaps the case of $\polp\star$. For this case, by the induction hypothesis, all
$\den{\polp^{(n)}}(a)$ are discrete probability distributions on packet sets, therefore vanish outside $\pPk$. By \cref{lem:convergence-on-cantor-open}, this is also true of the limit $\den{\polp\star}(a)$, as its value on $\pPk$ must be 1, therefore it is also a discrete distribution on packet sets.
\end{proof}

\begin{proof}[Proof of \cref{lem:ptwise-convergence}]
This follows directly from \cref{lem:convergence-on-cantor-open} and
\cref{prop:old-and-new-sem} by noticing that $\emph{any}$ set $A \subseteq \pPk$
is a finite boolean combination of basic open sets.
\end{proof}

\begin{proof}[Proof of \cref{thm:big-step-sound}]
It suffices to show the equality $\bden{\polp}_{ab}
= \den{\polp}(a)(\sset{b})$; the remaining claims then follow by
well-definedness of $\den{-}$. The equality is shown using
\cref{lem:ptwise-convergence} and a routine induction on $\polp$:

For $\polp = \pfalse, \ptrue, \match{f}{n}, \modify{f}{n}$ we have
\[
\den{\polp}(a)(\sset{b}) = \dirac{c}(\sset{b}) = \ind{b = c} = \bden{\polp}_{ab}
\]
for $c = \emptyset, a, \set{\pk \in a}{\pk.f = n}, \set{\pk[f:=n]}{\pk \in a}$,
respectively.

For $\pnot{\preda}$ we have,
\[\begin{array}{rl@{\quad}l@{\quad}l}
\bden{\pnot{\preda}}_{ab}
&= \ind{b \subseteq a} \cdot \bden{\preda}_{a,a-b}\\
&= \ind{b \subseteq a} \cdot \den{\preda}(a)(\sset{a-b}) &&\text{(IH)}\\
&= \ind{b \subseteq a} \cdot \ind{a-b=a \cap b_t}        &&\text{(\cref{lem:preds-boolean-algebra})}\\
&= \ind{b \subseteq a} \cdot \ind{a-b=a - (\Hist - b_t)}\\
&= \ind{b=a \cap (H-b_t)}\\
&= \den{\pnot{\preda}}(a)(b)                              &&\text{(\cref{lem:preds-boolean-algebra})}
\end{array}
\]

For $\punion{\polp}{\polq}$, letting $\mu = \den{\polp}(a)$ and
$\nu = \den{\polq}(a)$ we have
\[\begin{array}{rl@{\quad}l@{\quad}l}
\den{\punion{\polp}{\polq}}(a)(\sset{b})
&= (\mu \times \nu)(\set{(b_1,b_2)}{b_1 \cup b_2 = b})\\
&= \sum_{b_1,b_2} \ind{b_1 \cup b_2 = b} \cdot (\mu \times \nu)(\sset{(b_1,b_2)})
  && \\ 
&= \sum_{b_1,b_2} \ind{b_1 \cup b_2 = b} \cdot
  \mu(\sset{b_1}) \cdot \nu(\sset{b_2})\\
&= \sum_{b_1,b_2} \ind{b_1 \cup b_2 = b} \cdot
  \bden{p}_{ab_1} \cdot \bden{q}_{ab_2} &&\text{(IH)}\\
&= \bden{\punion{\polp}{\polq}}_{ab}
\end{array}\]
where we use in the second step that $b \subseteq \Pk$ is finite, thus
$\set{(b_1,b_2)}{b_1 \cup b_2 = b}$ is finite.

For $\pseq{\polp}{\polq}$, let $\mu = \den{\polp}(a)$ and $\nu_c = \den{\polq}(c)$
and recall that $\mu$ is a discrete distribution on $\pset\Pk$. Thus
\[\begin{array}{rl@{\quad}l@{\quad}l}
\den{\pseq{\polp}{\polq}}(a)(\sset{b})
&= \sum_{c \in \pset\Pk} \nu_c(\sset{b}) \cdot \mu(\sset{c})\\
&= \sum_{c \in \pset\Pk} \bden{\polq}_{c,b} \cdot \bden{\polp}_{a,c}\\
&= \bden{\pseq{\polp}{\polq}}_{ab}.
\end{array}
\]

For $\polp \oplus_r \polq$, the claim follows directly from the induction hypotheses.

Finally, for $\polp\star$, we know that
$\bden{\polp^{(n)}}_{ab} = \den{\polp^{(n)}}(a)(\sset{b})$
by induction hypothesis. The key to proving the claim is
\cref{lem:ptwise-convergence}, which allows us to take the limit
on both sides and deduce
\begin{equation*}
  \bden{\polp\star}_{ab} =
  \lim_{n \to \infty} \bden{\polp^{(n)}}_{ab} =
  \lim_{n \to \infty} \den{\polp^{(n)}}(a)(\sset{b}) =
  \den{\polp\star}(a)(\sset{b}). \qedhere
\end{equation*}
\end{proof}

\begin{proof}[Proof of \cref{lem:small-step-stochastic}]
  For arbitrary $a,b \subseteq \Pk$, we have
  \begin{align*}
    \sum_{a',b'}\sden{\polp}_{(a,b),(a',b')}
    &= \sum_{a',b'} \ind{b' = a \cup b} \cdot \bden{p}_{a,a'}\\
    &= \sum_{a'} \Big(\sum_{b'} \ind{b' = a \cup b} \Big) \cdot \bden{p}_{a,a'}\\
    &= \sum_{a'} \bden{p}_{a,a'} = 1
  \end{align*}
  where in the last step, we use that $\bden{\polp}$ is stochastic (\cref{thm:big-step-sound}).
\end{proof}

\begin{proof}[Proof of \cref{lem:star-mc-char}]
By induction on $n \geq 0$. For $n=0$, we have
\begin{align*}
\sum_{a'} \ind{b' = a' \cup b} \cdot \bden{\polp^{(n)}}_{a,a'}
&= \sum_{a'} \ind{b' = a' \cup b} \cdot \bden{\ptrue}_{a,a'}\\
&= \sum_{a'} \ind{b' = a' \cup b} \cdot \ind{a=a'}\\
&= \ind{b' = a \cup b}\\
&= \ind{b' = a \cup b} \cdot \sum_{a'} \bden{\polp}_{a,a'}\\
&= \sum_{a'} \sden{\polp}_{(a,b),(a',b')}
\end{align*}

In the induction step ($n > 0$),
\begin{align*}
&\phantom{={}} \sum_{a'} \ind{b' = a' \cup b} \cdot
  \bden{\polp^{(n)}}_{a,a'} \\
&= \sum_{a'} \ind{b' = a' \cup b} \cdot
  \bden{\punion{\ptrue}{\pseq{p}{p^{(n-1)}}}}_{a,a'} \\
&= \sum_{a'} \ind{b' = a' \cup b} \cdot
  \sum_{c} \ind{a' = a \cup c} \cdot \bden{\pseq{p}{p^{(n-1)}}}_{a,c} \\
&= \sum_{c} \left(\sum_{a'} \ind{b' = a' \cup b} \cdot \ind{a' = a \cup c} \right)
  \cdot \sum_k \bden{p}_{a,k} \cdot \bden{p^{(n-1)}}_{k,c} \\
&= \sum_{c,k} \ind{b' = a \cup c \cup b} \cdot
  \bden{\polp}_{a,k} \cdot \bden{p^{(n-1)}}_{k,c} \\
&= \sum_{k} \bden{\polp}_{a,k} \cdot
   \sum_{a'} \ind{b' = a' \cup (a \cup b)} \cdot \bden{p^{(n-1)}}_{k,a'} \\
&= \sum_{k} \bden{\polp}_{a,k} \cdot
   \sum_{a'} \sden{\polp}^{n}_{(k,a \cup b),(a',b')} \\
&= \sum_{a'}\sum_{k_1,k_2} \ind{k_2 = a \cup b} \cdot \bden{\polp}_{a,k_1} \cdot
   \sden{\polp}^{n}_{(k_1,k_2),(a',b')} \\
&= \sum_{a'}\sum_{k_1,k_2} \sden{\polp}_{(a,b)(k_1,k_2)} \cdot
   \sden{\polp}^{n}_{(k_1,k_2),(a',b')} \\
&= \sum_{a'} \sden{\polp}^{n+1}_{(a,b),(a',b')}
\qedhere
\end{align*}
\end{proof}

\begin{lemma}
\label{lem:inverse}
The matrix $X=I-Q$ in Equation \eqref{eq:inverse} of \Cref{sec:closed-form} is invertible.
\end{lemma}
\begin{proof*}
Let $S$ be a finite set of states, $\len S=n$, $M$ an $S\times S$ substochastic matrix ($M_{st}\geq 0$, $M\One\leq\One$). A state $s$ is \emph{defective} if $(M\One)_s < 1$. We say $M$ is \emph{stochastic} if $M\One=\One$, \emph{irreducible} if $(\sum_{i=0}^{n-1} M^i)_{st} > 0$ (that is, the support graph of $M$ is strongly connected), and \emph{aperiodic} if all entries of some power of $M$ are strictly positive.

We show that if $M$ is substochastic such that every state can reach a defective state via a path in the support graph, then the spectral radius of $M$ is strictly less than $1$. Intuitively, all weight in the system eventually drains out at the defective states.

Let $e_s$, $s\in S$, be the standard basis vectors. As a distribution, $e_s^T$ is the unit point mass on $s$. For $A\subs S$, let $e_A=\sum_{s\in A} e_s$. The $L_1$-norm of a substochastic vector is its total weight as a distribution. Multiplying on the right by $M$ never increases total weight, but will strictly decrease it if there is nonzero weight on a defective state. Since every state can reach a defective state, this must happen after $n$ steps, thus $\norm{e_s^TM^n}<1$. Let $c = \max_s \norm{e_s^TM^n}<1$.
For any $y=\sum_s a_s e_s$,
\begin{align*}
\norm{y^TM^n} &= \norm{(\sum_s a_s e_s)^TM^n}\\
&\leq \sum_s \len{a_s}\cdot\norm{e_s^TM^n}
\leq \sum_s \len{a_s}\cdot c
= c\cdot\norm{y^T}.
\end{align*}
Then $M^n$ is contractive in the $L_1$ norm, so $\len\lambda < 1$ for all eigenvalues $\lambda$.
Thus $I-M$ is invertible because $1$ is not an eigenvalue of $M$.
\end{proof*}

\begin{proof}[Proof of \cref{prop:SU-properties}]~
\begin{enumerate}[leftmargin=*]
\item It suffices to show that $USU = SU$. Suppose that
\[\prob{(a,b) \reaches{USU}_1 (a',b')} = p > 0. \]
It suffices to show that this implies \[
  \prob{(a,b) \reaches{SU}_1 (a',b')} = p.
\]
If $(a,b)$ is saturated, then we must have $(a',b') = (\emptyset,b)$ and \[
  \prob{(a,b) \reaches{USU}_1 (\emptyset,b)} = 1 =
    \prob{(a,b) \reaches{SU}_1 (\emptyset,b)}
\]
If $(a,b)$ is not saturated, then $(a,b) \reaches{U}_1 (a,b)$ with
probability $1$ and therefore \[
\prob{(a,b) \reaches{USU}_1 (a',b')} = \prob{(a,b) \reaches{SU}_1 (a',b')}
\]
\item Since $S$ and $U$ are stochastic, clearly $SU$ is a MC. Since $SU$ is finite
state, any state can reach an absorbing communication class.
(To see this, note that the reachability relation $\reaches{SU}$ induces a
partial order on the communication classes of $SU$. Its maximal elements are
necessarily absorbing, and they must exist because the state space is finite.)
It thus suffices to show that a state set $C \subseteq \pset\Pk \times \pset\Pk$
in $SU$ is an absorbing communication class iff
$C = \sset{(\emptyset,b)}$ for some $b\subseteq\Pk$.
  \begin{enumerate}[leftmargin=*]
  \item[``$\Leftarrow$'':]
    Observe that $\emptyset \reaches{B}_1 a'$ iff $a'=\emptyset$. Thus
    $(\emptyset,b) \reaches{S}_1 (a',b')$ iff $a'=\emptyset$ and $b'=b$,
    and likewise $(\emptyset,b) \reaches{U}_1 (a',b')$ iff
    $a'=\emptyset$ and $b'=b$.
    Thus $(\emptyset,b)$ is an absorbing state in $SU$ as required.
  \item[``$\Rightarrow$'':]
  First observe that by monotonicity of $SU$ (\cref{lem:monotone-chains}),
  we have $b=b'$ whenever
  $(a,b) \communicate{SU} (a',b')$; thus there exists a fixed $b_C$ such that
  $(a,b) \in C$ implies $b=b_C$.

  Now pick an arbitrary state $(a,b_C) \in C$. It suffices to show that
  $(a,b_C) \reaches{SU}(\emptyset,b_C)$, because that implies
  $(a,b_C) \communicate{SU}(\emptyset,b_C)$, which in turn implies $a=\emptyset$.
  But the choice of $(a,b_C) \in C$ was arbitrary, so that would mean
  $C = \sset{(\emptyset,b_C)}$ as claimed.

  To show that $(a,b_C) \reaches{SU} (\emptyset, b_C)$, pick arbitrary states
  such that \[
    (a,b_C) \reaches{S} (a',b') \reaches{U}_1 (a'',b'')
  \]
  and recall that this implies $(a,b_C) \reaches{SU} (a'',b'')$ by claim
  \eqref{prop:SU-properties:(SU)*=S*U}.
  Then $(a'',b'') \reaches{SU} (a,b_C)$ because $C$ is absorbing, and
  thus $b_C=b'=b''$ by monotonicity of $S$, $U$, and $SU$. But $(a',b')$ was
  chosen as an arbitrary state $S$-reachable from $(a,b_C)$, so
  $(a,b)$ and by transitivity $(a',b')$ must be saturated.
  Thus $a''=\emptyset$ by the definition of $U$. \qedhere
  \end{enumerate}
\end{enumerate}
\end{proof}

\begin{proof}[Proof of \cref{thm:closed-form}]
Using \cref{prop:SU-properties}.\ref{prop:SU-properties:(SU)*=S*U} in the second
step and equation \cref{eq:SU-limit-closed-form} in the last step,
\begin{align*}
\lim_{n\to\infty} \sum_{a'} S^n_{(a,b),(a',b')}
&= \lim_{n\to\infty} \sum_{a'} (S^nU)_{(a,b),(a',b')}\\
&= \lim_{n\to\infty} \sum_{a'} (SU)^n_{(a,b),(a',b')}\\
&= \sum_{a'} (SU)^\infty_{(a,b),(a',b')} = (SU)^\infty_{(a,b),(\emptyset,b')}
\end{align*}
$(SU)^\infty$ is computable because $S$ and $U$ are matrices over $\Q$ and hence
so is $(I-Q)^{-1}R$.
\end{proof}

\begin{corollary}
\label{cor:equiv-decidable}
For programs $p$ and $q$, it is decidable whether $\polp \equiv \polq$.
\end{corollary}

\begin{proof}[Proof of \cref{cor:equiv-decidable}]
Recall from \cref{cor:den-equal-iff-bden-equal} that it suffices to compute the
finite rational matrices $\bden{\polp}$ and $\bden{\polq}$ and check them for equality.
But \cref{thm:closed-form} together with \cref{prop:small-step-characterization}
gives us an effective mechanism to compute $\bden{-}$ in the case of Kleene star,
and $\bden{-}$ is straightforward to compute in all other cases. Summarizing
the full chain of equalities, we have:
\begin{equation*}
  \den{\polp\star}(a)(\sset{b}) = \bden{\polp\star}_{a,b}
  = \lim_{n\to\infty} \bden{\polp^{(n)}}_{a,b}
  = \lim_{n\to\infty} \sum_{a'} \sden{\polp}^n_{(a,\emptyset),(a',b)}
  = (SU)^\infty_{(a,\emptyset),(\emptyset,b)}
\end{equation*}
following from \cref{thm:big-step-sound}, Definition of $\bden{-}$,
\cref{prop:small-step-characterization}, and finally \cref{thm:closed-form}.
\end{proof}

\section{Handling Full \probnetkat: Obstacles and Challenges}
\label{sec:full}

History-free \probnetkat can describe sophisticated network routing schemes
under various failure models, and the program semantics can be computed exactly.
Performing quantitative reasoning in full \probnetkat appears significantly more
challenging. We illustrate some of the difficulties in deciding program
equivalence; recall that this is decidable for the history-free fragment
(\cref{cor:equiv-decidable}).

The main difference in the original \probnetkat language is an additional
primitive $\pdup$.  Intuitively, this command duplicates a packet $\pk \in \Pk$
and outputs the word $\pk\pk \in \Hist$, where $\Hist = \Pk^*$ is the set of
non-empty, finite sequences of packets. An element of $\Hist$ is called a
\emph{packet history}, representing a log of previous packet states. \probnetkat
policies may only modify the first (\emph{head}) packet of each history; $\pdup$
fixes the current head packet into the log by copying it.  In this way,
\probnetkat policies can compute distributions over the paths used to forward
packets, instead of just over the final output packets.

However, with $\pdup$, the semantics of \probnetkat becomes significantly more
complex.  Policies $\polp$ now transform sets of packet histories $a \in \pH$ to
distributions $\den{\polp}(a) \in \Dist(\pH)$. Since $\pH$ is uncountable, these
distributions are no longer guaranteed to be discrete, and formalizing the
semantics requires full-blown measure theory (see prior work for
details~\citep{probnetkat-scott}).

Without $\pdup$, policies operate on sets of packets $\pPk$; crucially, this is
a \emph{finite} set and we can represent each set with a single state in a
finite Markov chain.  With $\pdup$, policies operate on sets of packet histories
$\pH$.  Since this set is not finite---in fact, it is not even
countable---encoding each packet history as a state would give a Markov chain
with infinitely many states.  Procedures for deciding equivalence are not known
for such systems in general.

While in principle there could be a more compact representation of
general \probnetkat policies as finite Markov chains or other models
where equivalence is decidable, (\eg, weighted or probabilistic
automata~\citep{droste2009handbook} or quantitative variants of regular
expressions~\citep{alur2016regular}), we suspect that deciding
equivalence in the presence of $\dup$ may be intractable. As circumstantial
evidence, \probnetkat policies can simulate a probabilistic variant of multitape
automaton originally introduced by \citet{rabin1959finite}. We
specialize the definition here to two tapes, for simplicity, but \probnetkat
programs can encode any multitape automata with any fixed number of tapes.

\begin{definition}
\label{def:bi-auto}
Let $A$ be a finite alphabet. A \emph{probabilistic multitape
automaton} is defined by a tuple $(S, s_0, \rho, \tau)$ where $S$ is a
finite set of states; $s_0 \in S$ is the initial state; $\rho : S \to
(A \cup \{ \blank \})^2$ maps each state to a pair of letters $(u,
v)$, where either $u$ or $v$ may be a special blank character
$\blank$; and the transition function $\tau : S \to
\Dist(S)$ gives the probability of transitioning from one state to another.
\end{definition}

The semantics of an automaton can be defined as a probability measure
on the space $A^\infty \times A^\infty$, where $A^\infty$ is the set
of finite and (countably) infinite words over the alphabet
$A$. Roughly, these measures are fully determined by the probabilities
of producing any two finite prefixes of words $(w, w') \in A^* \times
A^*$.

Presenting the formal semantics would require more concepts from
measure theory and take us far afield, but the basic idea is simple to
describe. An infinite trace of a probabilistic multitape automaton over states
$s_0, s_1, s_2, \dots$ gives a sequence of pairs of (possibly blank)
letters:
\[
\rho(s_0), \rho(s_1), \rho(s_2) \dots
\]
By concatenating these pairs together and dropping all blank
characters, a trace induces two (finite or infinite) words over the
alphabet $A$. For example, the sequence,
\[
(a_0,\blank), (a_1, \blank), (\blank, a_2), \dots
\]
gives the words $a_0 a_1 \dots$ and $a_2 \dots$. Since the traces are
generated by the probabilistic transition function $\tau$, each
automaton gives rise to a probability measure over pairs of infinite words.

Probabilistic multitape automata can be encoded as \probnetkat
policies with $\pdup$. We sketch the idea here, deferring further
details to \Cref{app:bi-auto}. Suppose we are given an
automaton $(S, s_0, \rho, \tau)$. We build a \probnetkat policy over
packets with two fields, $\kw{st}$ and $\kw{id}$. The first field
$\kw{st}$ ranges over the states $S$ and the alphabet $A$, while the
second field $\kw{id}$ is either $1$ or $2$; we suppose the input set
has exactly two packets labeled with $\kw{id} = 1$ and $\kw{id} = 2$.
In a set of packet history, the two active packets have the same value
for $\kw{st} \in S$---this represents the current state in the
automaton. Past packets in the history have $\kw{st} \in A$,
representing the words produced so far; the first and second
components of the output are tracked by the histories with $\kw{id} =
1$ and $\kw{id} = 2$. We can encode the transition function $\tau$ as
a probabilistic choice in \probnetkat, updating the current state
$\kw{st}$ of all packets, and recording non-blank letters produced by
$\rho$ in the two components by applying $\pdup$ on packets with the
corresponding value of $\kw{id}$.

Intuitively, a set of packet histories generated by the resulting \probnetkat
term describes a pair of words generated by the original automaton. With a bit
more bookkeeping (see \cref{app:bi-auto}), we can show that two
probabilistic multitape automata are equivalent if and only if their encoded
\probnetkat policies are equivalent. Thus, deciding equivalence for \probnetkat
with $\pdup$ is harder than deciding equivalence for probabilistic multitape
automata; similar reductions have been considered before for showing
undecidability of related problems about \kat~\citep{K97c} and probabilistic
\netkat~\citep{Kahn17}.

Deciding equivalence between probabilistic multitape automata is a challenging
open problem. In the special case where only one word is generated (say, when
the second component produced is always blank), these automata are equivalent to
standard automata with $\epsilon$-transitions (\eg, see
\citet{mohri2000generic}). In this setting, non-productive steps can be
eliminated and the automata can be modeled as finite state Markov chains, where
equivalence is decidable. In our setting, however, steps producing blank letters
in one component may produce non-blank letters in the other. As a result, it is
not clear how to eliminate these steps and encode our automata as Markov chains.
Removing probabilities, it is known that equivalence between non-deterministic
multitape automata is undecidable~\citep{griffiths1968unsolvability}. Deciding
equivalence of deterministic multitape automata remained a challenging open
question for many years, until \citet{harju1991equivalence}
surprisingly settled the question positively;
\citet{worrell2013revisiting} later gave an alternative proof. If
equivalence of probabilistic multitape automata is undecidable, then equivalence
is undecidable for \probnetkat programs as well. However if equivalence turns
out to be decidable, the proof technique may shed light on how to decide
equivalence for the full \probnetkat language.

\section{Encoding 2-Generative Automata in Full \probnetkat}
\label{app:bi-auto}

To keep notation light, we describe our encoding in the special case where the
alphabet $A = \{ x, y \}$, there are four states $S = \{ s_1, s_2, s_3, s_4 \}$,
the initial state is $s_1$, and the output function $\rho$ is
\[
  \rho(s_1) = (x, \blank) \qquad
  \rho(s_2) = (y, \blank) \qquad
  \rho(s_3) = (\blank, x) \qquad
  \rho(s_4) = (\blank, y) .
\]
Encoding general automata is not much more complicated.  Let $\tau : S \to
\Dist(S)$ be a given transition function; we write $p_{i,j}$ for
$\tau(s_i)(s_j)$. We will build a \probnetkat policy simulating this automaton.
Packets have two fields, $\kw{st}$ and $\kw{id}$, where $\kw{st}$ ranges over $S
\cup A \cup \{ \bullet \}$ and $\kw{id}$ ranges over $\{ 1, 2 \}$. Define:
\[
  p \defeq \pseq{\pseq
  {\match{\kw{st}}{s_1}}
  {\pstar{\kw{loop}}}}
  {\modify{\kw{st}}{\bullet}}
\]
The initialization keeps packets that start in the initial state, while the
final command marks histories that have exited the loop by setting $\kw{st}$ to
be the special letter $\bullet$.

The main program $\kw{loop}$ first branches on the current state $\kw{st}$:
\[
  \kw{loop} \defeq \text{case } \begin{cases}
    \match{\kw{st}}{s_1} : \kw{state}1 \\
    \match{\kw{st}}{s_2} : \kw{state}2 \\
    \match{\kw{st}}{s_3} : \kw{state}3 \\
    \match{\kw{st}}{s_4} : \kw{state}4
  \end{cases}
\]
Then, the policy simulates the behavior from each state. For instance:
\begin{align*}
  \kw{state1} \defeq \bigoplus \begin{cases}
    \pseq{(
      \ite{\match{\kw{id}}{1}}
      {\pseq {\modify{\kw{st}}{x}} {\pdup}}
      {\ptrue})}
    {\modify{\kw{st}}{s_1}}
    \withp p_{1, 1} , \\
    \pseq{(
      \ite{\match{\kw{id}}{1}}
      {\pseq {\modify{\kw{st}}{y}} {\pdup}}
      {\ptrue})}
    {\modify{\kw{st}}{s_2}}
    \withp p_{1, 2} , \\
    \pseq{(
      \ite{\match{\kw{id}}{2}}
      {\pseq {\modify{\kw{st}}{x}} {\pdup}}
      {\ptrue})}
    {\modify{\kw{st}}{s_3}}
    \withp p_{1, 3} , \\
    \pseq{(
      \ite{\match{\kw{id}}{2}}
      {\pseq {\modify{\kw{st}}{y}} {\pdup}}
      {\ptrue})}
    {\modify{\kw{st}}{s_4}}
    \withp p_{1, 4}
  \end{cases}
\end{align*}
The policies $\kw{state2}, \kw{state3}, \kw{state4}$ are defined similarly.

Now, suppose we are given two probabilistic multitape automata $W, W'$ that differ only in
their transition functions. For simplicity, we will further assume that both
systems have strictly positive probability of generating a letter in either
component in finitely many steps from any state. Suppose they generate
distributions $\mu, \mu'$ respectively over pairs of infinite words $A^\omega
\times A^\omega$. Now, consider the encoded \probnetkat policies $p, p'$. We
argue that $\den{p} = \den{q}$ if and only if $\mu = \mu'$.\footnote{%
  We will not present the semantics of \probnetkat programs with $\pdup$ here;
  instead, the reader should consult earlier
  papers~\citep{probnetkat-cantor,probnetkat-scott} for the full
development.}

First, it can be shown that $\den{p} = \den{p'}$ if and only if $\den{p}(e) =
\den{p'}(e)$, where
\[
  e \defeq \{ \pi \pi \mid \pi \in \Pk \} .
\]
Let $\nu = \den{p}(e)$ and $\nu' = \den{p'}(e)$. The key connection between the
automata and the encoded policies is the following equality:
\begin{equation}
  \label{eq:bi-auto:pnk}
  \mu(S_{u, v}) = \nu(T_{u, v})
\end{equation}
for every pair of finite prefixes $u, v \in A^*$. In the automata distribution
on the left, $S_{u, v} \subseteq A^\omega \times A^\omega$ consists of all pairs
of infinite strings where $u$ is a prefix of the first component and $v$ is a
prefix of the second component. In the \probnetkat distribution on the right,
we first encode $u$ and $v$ as packet histories. For $i \in \{ 1, 2 \}$
representing the component and $w \in A^*$ a finite word, define the history
\[
  \h_i(w) \in \Hist \defeq
    (\kw{st} = \bullet, \kw{id} = i)
    , (\kw{st} = w[|w|], \kw{id} = i) 
    , \dots
    , (\kw{st} = w[1], \kw{id} = i) 
    , (\kw{st} = s_1, \kw{id} = i) .
\]
The letters of the word $w$ are encoded in reverse order because by convention,
the head/newest packet is written towards the left-most end of a packet history,
while the oldest packet is written towards the right-most end. For instance, the
final letter $w[|w|]$ is the most recent (\ie, the latest) letter produced by
the policy. Then, $T_{u, v}$ is the set of all history sets including $\h_1(u)$
and $\h_2(v)$:
\[
  T_{u, v} \defeq \{ a \in \pH \mid \h_1(u) \in a , \h_2(v) \in a \} .
\]
Now $\den{p} = \den{p'}$ implies $\mu = \mu'$, since \cref{eq:bi-auto:pnk} gives
\[
  \mu(S_{u, v}) = \mu'(S_{u, v}) .
\]
The reverse implication is a bit more delicate.  Again by \cref{eq:bi-auto:pnk},
we have
\[
  \nu(T_{u, v}) = \nu'(T_{u, v}) .
\]
We need to extend this equality to all cones, defined by packet histories $\h$:
\[
  B_{\h} \defeq \{ a \in \pH \mid \h \in a \} .
\]
This follows by expressing $B_{\h}$ as boolean combinations of $T_{u, v}$, and
observing that the encoded policy produces only sets of encoded histories, \ie,
where the most recent state $\kw{st}$ is set to $\bullet$ and the initial state
$\kw{st}$ is set to $s_1$.


\section{Background on Datacenter Topologies}
\label{app:topology}

Data center topologies typically organize the network fabric into
several levels of switches.

\paragraph*{FatTree.}%
A FatTree~\citep{al2008scalable} is perhaps the most common example of
a multi-level, multi-rooted tree topology. \cref{fig:fattree} shows a
3-level FatTree topology with 20 switches. The bottom
level, \emph{edge}, consists of top-of-rack (ToR) switches; each ToR
switch connects all the hosts within a rack (not shown in the figure).
These switches act as ingress and egress for intra-data center
traffic. The other two levels, \emph{aggregation} and
\emph{core}, redundantly connect the switches from the edge layer.

The redundant structure of a FatTree makes it possible to implement
fault-tolerant routing schemes that detect and automatically route
around failed links. For instance, consider routing from a source to a
destination along shortest paths---\eg, the green links in the figure
depict one possible path from $(s7)$ to $(s1)$. On the way from the
ToR to the core switch, there are multiple paths that could be used to
carry the traffic. Hence, if one of the links goes down, the switches
can route around the failure by simply choosing a different
path. Equal-cost multi-path (ECMP) routing is widely used---it
automatically chooses among the available paths while avoiding longer
paths that might increase latency.

However, after reaching a core switch, there is a \emph{unique}
shortest path down to the destination. Hence, ECMP no longer provides
any resilience if a switch fails in the aggregation layer (\cf the red
cross in \cref{fig:fattree}). A more sophisticated scheme could take a
longer (5-hop) detour going all the way to another edge switch, as
shown by the red lines in the figure. Unfortunately, such detours can
lead to increased latency and congestion.

\paragraph*{AB FatTree.}%
The long detours on the downward paths in FatTrees are dictated by the
symmetric wiring of aggregation and core switches. AB
FatTrees~\citep{liu2013f10} alleviate this by using two types of
subtrees, differing in their wiring to higher levels. 
\cref{fig:ften}(a) shows how to rewire a FatTree to make it an AB FatTree.
The two types of subtrees are as follows:
\begin{enumerate}[label={\textit{\roman*})}]
  \item{Type A:} switches depicted in blue and wired to core using dashed lines.
  \item{Type B:} switches depicted in red and wired to core using solid lines.
\end{enumerate}
Type A subtrees are wired in a way similar to FatTree, but Type B
subtrees differ in their connections to core switches. In our
diagrams, each aggregation switch in a Type A subtree is wired to
adjacent core switches, while each aggregation switch in a Type B
subtree is wired to core switches in a staggered manner. (See the
original paper by \citet{liu2013f10} for the general construction.)

This slight change in wiring enables much shorter detours around
failures in the downward direction. Consider again routing from source
($s7$) to destination ($s1$).  As before, we have multiple options
going upwards when following shortest paths (\eg, the one depicted in
green), as well as a unique downward path. But unlike FatTree, if the
aggregation switch on the downward path fails, there is a short
detour, as shown in blue.
%
%
This path exists because the core switch, which needs to re-route
traffic, is connected to aggregation switches of both types of
subtrees. More generally, aggregation switches of the same type as the
failed switch provide a 5-hop detour; but aggregation switches of the
opposite type provide an efficient 3-hop detour.